\RequirePackage{fix-cm}
\documentclass[draft]{agujournal2019}
\usepackage{url} 
\usepackage{lineno}
\usepackage[finalnew]{trackchanges}
\usepackage{soul}

\usepackage{natbib}
\usepackage{amsmath}
\usepackage{amssymb}

\journalname{JGR: Solid Earth}

\begin{document}

%
%


\title{Differentiating frictionally locked asperities from kinematically coupled zones}

%
%




\authors{Dye SK Sato\affil{1}, Takane Hori\affil{1}, Yukitoshi Fukahata\affil{2}}

\affiliation{1}{Japan Agency for Marine-Earth Science and Technology, Kanagawa, Japan}
\affiliation{2}{Disaster Prevention Research Institute, Kyoto University, Japan}





\correspondingauthor{Daisuke Sato}{daisukes@jamstec.go.jp}



\begin{keypoints}
\item A formula to differentiate plate locking from plate coupling is derived, which enables estimation of locked asperities
\item Unified representations of locking and unlocking consist in pre-yield and post-yield phases of the failure criterion upon frictional slip
\item Asperities are detected around seafloor basins, fringed by slip zones of slow earthquakes kinematically coupled but frictionally unlocked
\end{keypoints}

%
%

%
%


\begin{abstract}
Seismogenic areas on plate-boundary faults resist slipping until earthquakes begin. The delay in slip relative to plate motion, termed slip deficit, represents plate coupling as an interseismic proxy of seismic potential. However, when a part of a frictional interface sticks together (locked), the unlocked sliding surroundings are braked and slowed (coupled), causing coupled zones always wider than the locked zones that rupture during earthquakes. This study investigates the frictional physics that locked and unlocked zones should observe, laying the foundation for inferring frictionally locked segments, known as asperities in fault mechanics. Various friction laws are shown to have a unified representation of locking. (I) Locking means the pre-yield phase, where the fault interface does not slip, and unlocking means the post-yield phase, where stress on the interface equals strength. (II) For intersesismic periods, while locking still denotes constant slip, unlocking signifies quasi-steady creeping of constant stress. Locking inversion, a variant of conventional coupling inversion that incorporates this unified frictional physics, estimates the distribution of locking, determining slip and stress distributions consequently. We solve the locking inversion by a method that distributes circular asperities on unlocked interfaces. By applying this method to on-/off-shore GNSS data in southwestern Japan, we detect five primary locked segments along the Nankai subduction zone. Those segments accord with slip zones of historical megathrust earthquakes correlated with seafloor basins. Estimated locked zones avoid the occurrence zones of deep slow earthquakes, reproducing the hypothesis that the areas hosting slow earthquakes are normally, in interseismic timescales, coupled but unlocked.
\end{abstract}

\section*{Plain Language Summary}
Interplate earthquakes slide the stationary contacts of neighboring plates. Approaching earthquakes may therefore occur in coupled zones of the interface, where the sliding speed is slower than the plate motion. However, locked, unmoving contacts pin the surroundings, making the coupled zones significantly wider than the locked zones that become the source of the earthquake. We examine the physics of friction that causes plate coupling, to refine the concepts of plate locking and plate unlocking. These concepts stand on a principle that many friction laws follow: locking refers to an interface without sliding, and unlocking refers to an interface at a force threshold. Considering a long period between earthquakes, we show that an unlocked zone nearly equals an interface without force change. This constraint of constant slip or force is the unified law that governs coupling. By combining it with the estimation of coupling, we can estimate locked zones, which are the cause of plate coupling and force loading. To test the derived formula, we apply it to geodetic data around the Nankai subduction zone in Japan. Our estimates reproduce existing hypotheses that hazardous, huge locked zones correlate with offshore basins surrounded by slow-earthquake areas inside creeping zones.

%
%

%


%
%
%
%

\section{Introduction}
Seismogenic zones in interseismic periods store seismic moment to be released. 
The accumulated moment, known as slip deficit, is regarded as the coupling of adjoining plates and serves as a proxy of seismogenic zones on plate boundaries~\citep{kanamori1971great,savage1983dislocation}. 
According to kinematic slip deficit inversions, also called coupling inversions, which estimate slip deficit from surface displacement data, highly coupled zones correlate well with coseismic slip zones~\citep{scholz2012seismic}.

Meanwhile, fault rupture is a stick-slip phenomenon in which \remove{a stress-loaded stationary zone (a locked zone)}\add{a stationary zone on a stress-loaded interface, termed a locked zone,} slips when \add{the shear stress acting on }it reaches a threshold stress~\citep{reid1910mechanism}. 
Then, when comparing locking and coupling, contrasting concepts of frictional failure and moment release, 
it becomes a problem that the coupled zone is always wider than the locked zone~\citep{ruff1983seismic,wang1995coupling,herman2018accumulation}. 
In kinematic terms solely relying on the slip rates $V$ on a plate boundary, 
locking refers to zero slip rate (full coupling, $V=0$), whereas the surrounding unlocked zones produce finite slip rates $V$ significantly lower than the plate convergence rate $V_{\rm pl}$ (partial coupling, $0 < V < V_{\rm pl}$)~\citep{wang1995coupling}. 
In short, locked zones brake surrounding unlocked zones, complicating the interpretation of coupling~\citep{wang2004coupling,burgmann2005interseismic}. 

This longstanding issue of coupling-locking \remove{semantics}\add{differentiation have affected several problems in understanding and assessing earthquakes. Examples include the quantification of coseismic slip zones, which are narrower than the coupled zones due to the presence of aseismic afterslip that accompanies coseismic slip (Saito \& Noda, 2022). It also} extends to the slow earthquake literature, as it has been suggested that steadily highly coupled zones (i.e., presumably locked zones) correspond to the source regions of paleoseismic megathrust earthquakes, while moderately coupled zones (i.e., presumably close to locked zones) correspond to the slip regions of slow earthquakes~\citep{baba2020slow} in a long-term sense, although slip zones of slow earthquakes can vary
coupling ratios during own recurrence intervals~\citep{bartlow2020long,wallace2020slow}. 
\add{Those present-day problems stagnate in disambiguating the locked zone subject to coseismic failure from the coupled zone subject to any non-interseismic strain release.}

The coupling-locking differentiation problem is twofold; one appears in interpretation, thus conceptual, and the other matters in quantification, thus practical. 
Regarding the concepts, it is common for coupling to be equated to locking in result interpretations. 
\citet{wang2004coupling} criticize this convention, working on the classification of often-confused mechanical concepts (sliding, stressing, locking, and strength), and emphasize that coupling is nothing more than information on sliding. 
Regarding the practices, 
even knowing that coupling is different from locking, plate locking is often discussed in terms of the coupling ratio (full coupling or partial coupling, etc.), largely due to the lack of established indicators of locking. 
However, the spatial variation in coupling is blurred by inversion errors and biases, making it hard to extract locked zones of exact $V=0$ from highly coupled zones based on kinematic coupling estimates alone~\citep{burgmann2005interseismic}. 

Therefore, pioneering research is towards directly inferring the mechanics that results in plate coupling, instead of interpreting it from inferred coupling. 
Several mechanical indicators other than (kinematic) coupling have been proposed, now collectively referred to as mechanical coupling~\citep{herman2020locating,saito2022mechanically}.

Mechanical coupling inference is the practice of coupling semantics that untangles mechanical concepts lumped together with kinematic coupling~\citep{wang2004coupling}. Preexisting mechanical couplings are classified into two types: stressing (force), linear transformation of the slip deficit, and locking (friction), defined in the sense of Amontons-Coulomb friction, presuming instant transition between static and dynamic frictions.  
A distinction between coupling (slip), stressing (force), and locking (friction) has been clear since \citet{wang2004coupling}, and our terminology follows theirs. In this paper, we avoid using the polysemantic `mechanical coupling' for clarity. `Kinematic coupling' is consistently called `coupling' hereafter.

Stressing (stressing rate) represents the rate of stress accumulation due to coupling (slip deficit). Stressing inversion imposes a priori constraints on stress loading, whereas conventional coupling inversion imposes a priori constraints on slip. 
Stressing inversion is a simple linear transform of coupling inversion that converts slip to stress but can detect stress-loaded regions closely related to the locked zone~\citep{noda2021energy,saito2022mechanically}. 
Constraints on stressing help obtain physically reasonable estimates of coupling~\citep{lindsey2021slip}.

Locking is defined in the sense of Amontons-Coulomb (or ``static-dynamic'') friction, thus far. In Amontons-Coulomb friction, the static-frictional region of zero sliding is locked, and the dynamic-frictional region of constant stress is unlocked~\citep{burgmann2005interseismic,funning2007asperities,johnson2010new,herman2020locating}. 
This physical constraint sets a nonlinear problem
to calculate the coupling field under the given boundary conditions that impose zero slip rate inside locked zones and zero stressing rate outside, according to the binary field to express locking. The coupling field calculated as a functional of the locking field in turn gives the surface displacement. Locking inversion estimates the locked zone by performing an inverse analysis of such a two-stage forward model. 

The above survey on coupling-locking differentiation allows us to recognize a crucial piece of information missing: How to relate those indicators to {\it true} locking? \add{As recapitulated, when coupling and stressing, or slip deficit and traction are sought, the analysis is to invert observed on-ground motion to unknown on-fault motion}~\citep{savage1983dislocation}\add{. In this type of analysis, locking is full coupling ($V=0$), and unlocking is the remaining ($V>0$). 
This kinematic definition of locking has suffered from vulnerability to inference errors of coupling}~\citep{burgmann2005interseismic}\add{ because slip deficits gradually decrease from full coupling}~\citep{wang1995coupling}.
\add{The locking assessment then requires a criterion to resolve the kinematic absurd to distinguish $V=0$ from $V=+0$, typified by coupling cutoff and the Amontons-Coulomb friction, in return for the apparent criterion dependence}~\citep{burgmann2005interseismic,funning2007asperities}. 
\add{As all proposers are aware, any measures of kinematic and mechanical couplings do not truly replicate the locking thus far}~\citep{ruff1983seismic,burgmann2005interseismic,saito2022mechanically}.

We use the word {\it true} in the sense of inference, which refers to an ideal estimate available in the limit of complete observations (data) with complete forward models (observation
equations)~\citep{yagi2011introduction}. While limitations of observation~\citep{yokota2016seafloor} and 
Green's function errors~\citep{yagi2011introduction} have been closely examined, the model errors of plate interface rheology wait for scrutiny.
One very close indicator, a reasonable model, of the true locking will be the above-mentioned ``locking'' defined in the Amontons-Coulomb sense. 
Yet, Amontons-Coulomb locking is not the only possibility. It becomes a problem when comparing Amontons-Coulomb locking to recent findings, including slow earthquakes, where various physical interpretations have been tried based on countless friction laws. Examples include slow slip events modeled by rate-and-state friction with velocity cutoff~\citep[e.g.,][]{shibazaki2003physical}, fluid-induced tremors~\citep[e.g.,][]{yamashita2011dynamic}, and tremors at depth as a
semi-brittle failure inside a brittle-ductile transition zone~\citep{ando2012propagation}. 
Comparisons between results using different friction laws~\citep{sherrill2024locating} provide a valuable guess of the model errors of assumed specific laws. However, no one knows the true physics of plate boundaries, so the model error quantification of plate locking has largely been unaddressed. 
The relationship between Amontons-Coulomb locking, other locking indicators depending on other laws, and true locking is, thus, not yet clear.

This study explores the physics that specifies locking and unlocking, based solely on the premise that fault motion is frictional slip.
Our study begins by engaging on a failure criterion universal to frictional failure, known as the yield criterion, 
including a subtle refinement to frictional constitutive law since rate-and-state friction (\S\ref{sec:2}).
We will show that the complementarity of friction (eq.~\ref{eq:complementarity_strengthexcess}) plays a key role in characterizing unlocking, which is the point lacked in the kinematics. 
We then prove that a simple constraint results from the yield criterion during interseismic phases (eq.~\ref{eq:complementarity_stressingsliprates}). For interseismic periods, any laws
result in the same constraint,
and so we can treat the Amontons-Coulomb locking as true locking. 
\add{Once the governing equation of locking is identified, the relationship between kinematic and mechanical couplings is naturally understood} (\S\ref{subsec:semantics}).
\add{Plate locking} (eq.~\ref{eq:locking_constraint}) \add{is shown to be the boundary condition that determines slip deficit} (coupling, eq.~\ref{eq:sfromsd}) \add{and stress loading} (stressing, eq.~\ref{eq:slip2traction}). 

The above finding forces us to address a practical issue that the inversion of the Amontons-Coulomb locking is a highly nonlinear inference that produces a multimodal (multi-peaked) probability (\S\ref{sec:transdim}). 
For robust estimation of locking, we construct a transdimensional locking-inversion scheme, in which
the number of model parameters is optimized, along the lines of the concept of locked-zone segments, known as asperities in fault mechanics. 
Last, we apply our method to the Nankai subduction zone in southwestern Japan (\S\ref{sec:Application}) to examine the workings of the proposed scheme in actual data analysis. 
Through the validation of our method and the derived formula of locking, 
we report that the locked zones estimated from geodetically observed data accord with the characteristics of the known regular and slow earthquake activity (\S\ref{sec:Discussion}).

\section{Observation Equation in Locking Inversion}
\label{sec:2}
This section is our examination of the observation equation in locking inversion, where
we reconcile the hypothetical Amontons-Coulomb locking with the true locking of plate boundaries. 
We start by looking at the formulation of slip deficit inversion (\S\ref{subsec:slipdeficitinv}) 
since the locking inversion is a variant of the slip deficit inversions that imposes physical constraints to link coupling and locking. To better understand locking inversions, 
we will explain the assumption of quasi-stationarity, often used in slip deficit inversions.
This assumption states that slip acceleration is negligible over long periods of time in the inter-seismic period, offering a principle of locking inversions. 
For the same purpose, we also emphasize that the conventional slip deficit inversion assumes that complete uncoupling causes negligible interseismic deformation only. 
Next, we examine the original frictional definition of locking in accordance with the yield criterion universal among friction laws (\S\ref{subsec:complementarity}). 
Then, we reduce such various friction-law-dependent representations of locking to a single universal friction-law-independent formula by the approximation of quasi-stationarity, which is, as mentioned earlier, equivalent to the Amontons-Coulomb locking (\S\ref{subsec:derivinglockinginversion}). 
Last, we summarize the derived results in terms of coupling semantics (\S\ref{subsec:semantics}). 
We will see the spatial relationship between kinematic coupling and mechanical couplings (coupling, stressing, and locking). 
In this section, we shall clarify that while the slip deficit inversion is the inversion of the so-called dislocation problem, the locking inversion is the inversion of a crack problem. 

\subsection{Slip deficit inversion as inverse dislocation problem}
\label{subsec:slipdeficitinv}

Suppose that we observe the crustal deformation rate at points $i=1,...,N$ and that from them, we extract the deformation-rate components $\dot u_i$ associated with relative motions of a plate boundary $\Gamma$. 
The slip deficit inversion~\citep{savage1983dislocation} estimates the crustal-deformation-inducing slip $s_{\rm d}$ at the plate boundary from $\dot u_i$. To clearly write down the assumption of locking inversions (later in \S\ref{subsec:derivinglockinginversion}), 
we account for the fact that $\dot u_i$ depends on the observation period $t\in(0,\Delta t)$ and distinguish time-varying $\dot u_i$ from its long-term trend $d_i$, where $\Delta t$ denotes the observation duration.

If the deformation of interest is limited to that of the hangingwall (e.g., all observation points are located on the upper plate of the subduction zone), the forward model of deformation is a simple linear form. 
The deformation of the hangingwall is due to the internal forces in the hangingwall
and footwall, and therefore the momentum and angular momentum are conserved; that conservative force is generally written by seismic moment $\mathcal M$~\citep{backus1976momenti,backus1976momentii}: 
\begin{equation}
    \dot u_i(t)=\int_\Gamma d\Sigma(\boldsymbol\xi) G^{(\mathcal M)}_i\dot {\mathcal M}(\boldsymbol\xi,t)+e_i(t),
    \label{eq:originalObsEq}
\end{equation}
where $G^{(\mathcal M)}$ denotes Green's function that relates the moment rate $\dot {\mathcal M}$ and the deformation rate, and $e_i$ represents the error term. We omit to write down the vectorial
nature of $u_i$ and the tensorial nature of $\dot {\mathcal M}$.
When observations exist also on the footwall, the same holds after correcting the rigid-body translation of the two plates.

Equation~(\ref{eq:originalObsEq}) shows that the slip deficit inversion is an inverse problem of the dislocation problem, which estimates the interseismic moment accumulation on the plate interface. 
At the same time, the stress accumulation rate on the plate boundary (stressing rate) $\dot T$ is also expressed in a linear form: 
 \begin{equation}
    \dot T({\bf x},t)=\int_\Gamma d\Sigma(\boldsymbol\xi) K^{(\mathcal M)}({\bf x},\boldsymbol\xi) \dot {\mathcal M}(\boldsymbol\xi,t),
    \label{eq:originalObsEq4traction}
\end{equation}
where $K^{(\mathcal M)}$ denotes Green's function that relates $\dot {\mathcal M}$ to plate traction rate. We omit to write down $T$ at each point vectorially. 
The basis of the later-introduced locking inversion is the feasibility of tracking stress loads (stressing) during moment accumulation (coupling), regardless of the moment's origin. 
Therefore, eq.~(\ref{eq:originalObsEq4traction}) is an important equality, which holds regardless of the controversial interpretation of $\mathcal M$ outlined at the last of this subsection. 

\begin{figure*}
   \includegraphics[width=150mm]{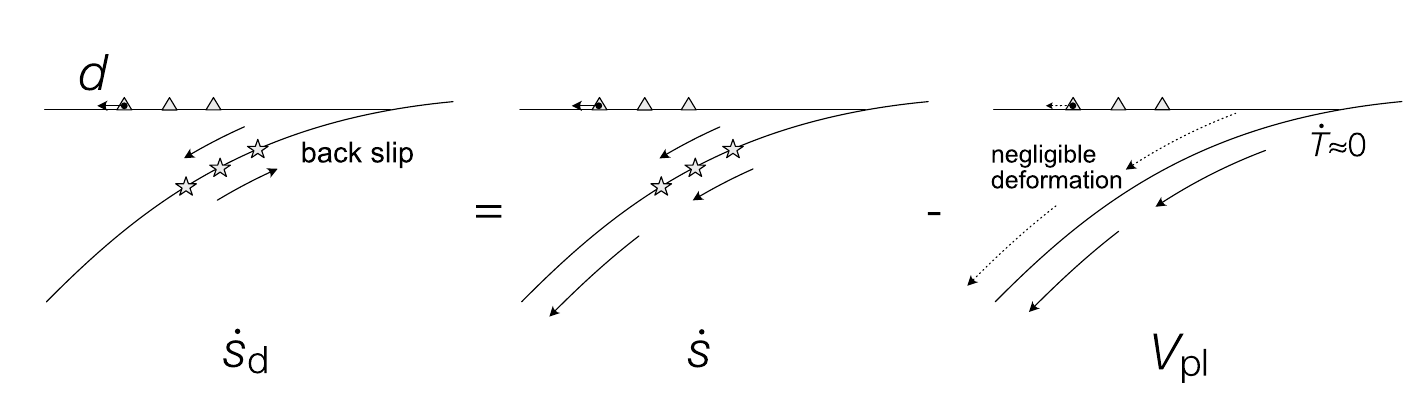}
  \caption{
  Relationship among the slip deficit rate $\dot s_{\rm d}$, slip rate $\dot s$, and long-term subduction rate $V_{\rm pl}$, shown in the inertial coordinate of the hangingwall. 
  The slip is decomposed into long-term part $V_{\rm pl}$ and the residual $\dot s_{\rm d}$. 
  Assuming that crustal deformation from $V_{\rm pl}$ (dotted lines in the figure) is negligible, the slip deficit inversion ascribes observed surface deformation to $\dot s_{\rm d}$. This approximation corresponds to identifying the subduction at $V_{\rm pl}$ as an approximately traction-free solution.
}
  \label{fig:sdsVpl}
\end{figure*}

Then, we construct the slip deficit inversion in an ordinary way (Fig.~\ref{fig:sdsVpl}).
The conventional slip deficit inversion decomposes
slip rate $\dot s$ of the plate interface (the relative velocity of plate boundaries) into 
the relative rigid-body velocity $V_{\rm pl}$ and the residual $\dot s_{\rm d}$, 
\begin{equation}
    \dot s= V_{\rm pl}-\dot s_{\rm d},
    \label{eq:sfromsd}
\end{equation}
and assumes the crustal deformation due to $V_{\rm pl}$ is negligible. 
That is, large parts of surface deformations (deviations from the rigid-body plate motion) are assumed to come from the slip deficit $\dot s_{\rm d}$: 
\begin{equation}
    \dot{\mathcal M} \simeq -C\nu \dot s_{\rm d},
    \label{eq:Moment2slipdeficit}
\end{equation}
where $C$ denotes the stiffness tensor, and $\nu$ denotes the plate normal. 
Following convention, the direction of $\dot s_{\rm d}$ is set to the opposite from that of the subduction (back slip). 
This approximation of eq.~(\ref{eq:Moment2slipdeficit})  attributes the drag force of the continental plate to the deficit in the subducting motion of the oceanic plate delayed from the relative rigid-body motion of the two plates.

After the approximation of eq.~(\ref{eq:Moment2slipdeficit}), the deformation rate $\dot u_i$ is given by a linear function of $\dot s_{\rm d}$;
from eqs.~(\ref{eq:originalObsEq}) and (\ref{eq:Moment2slipdeficit}), 
\begin{equation}
    \dot u_i(t)=\int_\Gamma d\Sigma(\boldsymbol\xi) G_i\dot s_{\rm d}(\boldsymbol\xi,t)+e_i(t),
    \label{eq:sdi_obseq}
\end{equation}
where $G(=-G^{(\mathcal M)} C\nu)$ denotes Green's function that relates $\dot s_{\rm d}$ to surface displacement rates. The error term $e_i$ in eq.~(\ref{eq:originalObsEq}) is redefined in eq.~(\ref{eq:sdi_obseq}) to include the approximation error of eq.~(\ref{eq:Moment2slipdeficit}).

Similarly, from eqs.~(\ref{eq:originalObsEq4traction}) and (\ref{eq:Moment2slipdeficit}),
\begin{equation}
    \dot T\simeq \int_\Gamma d\Sigma K\dot s_{\rm d},
    \label{eq:slip2traction}
\end{equation}
where $K(=-K^{(\mathcal M)} C\nu)$ denotes the traction Green's function on the plate boundary. Equation~(\ref{eq:slip2traction}) states that no stress loading ($\dot T=0$) approximately means no coupling ($\dot s_{\rm d}=0$).

Moreover, it is common to fit $\dot u_i$ by a linear trend over the analysis period $t\in(0,\Delta t)$: 
\begin{equation}
\dot u_i(t)\simeq d_i:=\frac 1{\Delta t}\int^{\Delta t}_0 dt^\prime \dot u_i(t^\prime),    
\end{equation}
which reduces eq.~(\ref{eq:sdi_obseq}) to
\begin{equation}
    d_i=\int_\Gamma d\Sigma(\boldsymbol\xi) G_i\dot s_{\rm d}(\boldsymbol\xi)+e_i,
\end{equation}
and
\begin{equation}
    \ddot s_{\rm d}\simeq 0.
    \label{eq:quasistationary}
\end{equation}
Equation~(\ref{eq:quasistationary}) represents the  approximation of quasi-stationarity that indicates the smallness of the time variation in $\dot s_{\rm d}$, which becomes essential to derive the locking inversion. 
The approximation error of quasi-stationarity is included in the error term $e_i$.

The error term is approximated by a Gaussian in many studies, which we follow: 
\begin{equation}
    {\bf e}\sim \mathcal N({\bf 0},{\bf C}_{\bf e}).
    \label{eq:GaussianError}
\end{equation}
where ${\bf e}$ is a vector notation of $e_i$, and ${\bf C}_{\bf e}$ denotes its covariance. 
The error quantification is not the scope of this paper, but we later show in \ref{subsec:appresult} that the error term $e_i$ is attributed partly to Green's function errors~\citep{yagi2011introduction}, rather than to observation errors alone. 

We end this subsection by outlining ongoing debates on the approximation of eq.~(\ref{eq:Moment2slipdeficit}) that attributes interseismic surface deformations to slip deficits. 
The central question in \S\ref{subsec:slipdeficitinv} is the estimation of the boundary motion between continental and oceanic plates, the bulks of which pass each other at the rigid-body velocity $V_{\rm pl}$. Such can be formulated as one branch of inverse dislocation problems that estimate the distribution of on-fault slip, under the remote boundary condition imposing the velocity difference $V_{\rm pl}$ at infinity. 
Then, in solving this problem, eq.~(\ref{eq:Moment2slipdeficit}) neglects the crustal deformation due to the slip at $V_{\rm pl}$, expecting no stress loading if no coupling. 
However, it is an approximation because the constant rate subduction is not traction-free for non-planar plate boundaries with finite curvature~\citep{savage1983dislocation,hashimoto20043,hashimoto20063,fukahata2016deformation,romanet2024mechanics}. That is, when an oceanic plate moves at a convergent plate speed, the upper plate also deforms. 
More fundamentally, the accumulated stress due to long-term subduction is relieved by off-fault inelastic deformations of brittle or
ductile rheology~\citep{searle1987closing}, so a part of the deformation is accumulated but never restored elastically. 
Considering such seismically unreleased portions of coupling, the coupling ratio may not be a good proxy of the seismic potential but rather its upper bound. 


\subsection{Complementarity of slip rate and strength excess on a frictional interface}
\label{subsec:complementarity}

Conventional coupling inversions treat only the slip deficit $s_{\rm d}$, or equivalently, only the coupling ratio $\dot s_{\rm d}/V_{\rm pl}$. 
When estimating locking as well as coupling, modern geodetic inversions premise the Amontons-Coulomb friction as mentioned earlier. The aim of this study is to evaluate the model bias due to such use of a specific friction law. For this purpose, we must not rely on functional forms of specific laws because the true law of fault motions is never known. Thus, we attend to the very universal, a priori definition of locking, which derives from the yield criterion various friction laws observe.

In terms of mechanics, frictional sliding is one form of fracture~\citep{scholz2019mechanics}. 
A fairly large number of friction laws describe the onset conditions of frictional sliding by failure criteria. 
Those failure criteria are almost always included in the yield criterion below~\citep{smai2017study}; 
under the yield criterion of frictional failure,
the shear stress $T$ and slip rate $\dot s$ on the interface with frictional strength $\Phi$ obey the following branched condition (a mixed boundary condition):
\begin{equation}    
    \begin{aligned}        
T<\Phi &\cap \dot s=0 \hspace{10pt}{\rm (locking)}
\\
T=\Phi &\cap \dot s>0 \hspace{10pt}{\rm (unlocking)}
    \end{aligned}
    \label{eq:yieldinglaw}
\end{equation}
In this paper, we do not distinguish traction and stress carefully. 
The top and bottom of eq.~(\ref{eq:yieldinglaw}) correspond to pre- and post-yield phases, respectively. 
The top of eq.~(\ref{eq:yieldinglaw}) states that the sliding starts when the stress $T$ on a crack face reaches the threshold stress, which is the frictional strength $\Phi$. 
The bottom of eq.~(\ref{eq:yieldinglaw}) indicates that strength refers not only to the threshold stress but also to the stress values of the post-yield interface.
Many friction laws
follow eq.~(\ref{eq:yieldinglaw}). Examples include Amontons-Coulomb friction and slip-weakening friction. Dynamic rupture simulations, including the models incorporating rate-weakening friction for fast
sliding, are usually based on eq.~(\ref{eq:yieldinglaw})~\citep[e.g.,][]{andrews1976rupture,cochard1994dynamic,harris2009scec}.

The rate- and state-dependent friction law~\citep[RSF law;][]{dieterich1979modeling}, commonly used in earthquake simulations, is a refinement of the yield criterion. 
\citet{dieterich1979modeling} discovered instantaneous stress change responding to slip-rate variations, termed the direct effect.
As elucidated in \citet{nakatani2001conceptual},
the direct effect is the signification of the constitutive law that relates the stress and slip rate: 
\begin{equation}
    T = A \ln (\dot s/V_*)+ \Phi,
\end{equation}
where $A$ represents the magnitude of the direct effect. $V_*$ is an arbitrary constant to represent the reference slip rate, and the RSF shows that $\Phi$ slightly depends on one's choice of $V_*$. $V_*$ is ordinarily set at the velocity of loading, now $V_*=V_{\rm pl}$. 
The $\Phi$ variations (state effects) in the RSF are often parametrized as $B\ln(\theta/\theta_*)$ with a conventional representation $\theta$ of the state variable, and two forms of the frictional state $\theta$ and $\Phi$ have one-to-one correspondence. 
This $\Phi$-notation signifies the crux of the RSF paradigm~\citep{nakatani2001conceptual}: the ``state'' in the rate-and-state friction is after all the strength $\Phi$, and the rate($V$)-and-state($\Phi$) description of
the frictional stress ($\tau$) is thus the rate-theoretic refinement of the yield criterion
(eq.~\ref{eq:yieldinglaw}) that switches stick and slip based on the stress ($\tau$) relative to the state of the interface ($\Phi$): 
\begin{equation}
    \dot s=V_{\rm pl}e^{(T-\Phi)/A}.
\end{equation}
This reading of the RSF as a flow law is consistent with Peierls thermal activation mechanisms of stick-slip phenomena~\citep{heslot1994creep}, investigated by experiments of \citet{nakatani2001conceptual}, 
and roughly consistent with the adhesion theory of friction relating the strength and real contact area, as shown by \citet{nagata2008monitoring} and \citet{nagata2014high} from acoustic and optical monitoring of frictional strength. 
Although the functional form $A\ln V$ of the direct effect is still under debate~\citep[e.g.,][]{barbot2019modulation}, the constitutive law of friction is the rheology of the yield criterion. 
The $A$ value is almost two digits smaller than fault normal stress, 
and thus the slip rate is negligible if $T$ is significantly smaller than $\Phi$, while a finite slip rate appears
if $T$ is close to $\Phi$; as a lowest order approximation of the RSF constitutive law with respect to $(T-\Phi)/A$, 
\begin{equation}    
    \begin{aligned}        
\Phi-T \gg A &\cap \dot s\ll V_{\rm pl} \hspace{10pt}{\rm (locking)}
\\
\Phi-T \lesssim A &\cap \dot s\gtrsim V_{\rm pl} \hspace{10pt}{\rm (unlocking)}
    \end{aligned}
    \label{eq:yieldinglaw_RSF}
\end{equation}
Equation~(\ref{eq:yieldinglaw_RSF}) refines the discontinuous approximation of eq.~(\ref{eq:yieldinglaw}) so that 
the moment of yielding ($\Phi\simeq T$) with negligible slip rates ($V/V_{\rm pl}\ll1$) can be tracked continuously~\citep{nakatani2001conceptual}. 
Conversely, when excluding the moment of yielding, even the RSF law is approximately within the realm of the classical yield criterion (eq.~\ref{eq:yieldinglaw}).

Friction laws established so far generally apply to the yield criterion (eq.~\ref{eq:yieldinglaw}) as above. 
Thus, it is worth noting that in eq.~(\ref{eq:yieldinglaw}), either strength excess $\Phi-T$, strength $\Phi$ relative to stress $T$, or the slip rate $\dot s$ is always zero~\citep{smai2017study}:
\begin{equation}
    (\Phi-T)\dot s= 0.
    \label{eq:complementarity_strengthexcess}
\end{equation}
In the literature of optimization theory, two variables are said to be complementary when the product of the two variables is always zero.
Equation~(\ref{eq:complementarity_strengthexcess}) states that the strength excess $\Phi-T$ and the slip rate $\dot s$ are complementary. 
Complementarity-based crack modeling can be found in solid and structural mechanics~\citep{bolzon2017complementarity} and geophysical applications~\citep{mutlu2008patterns,smai2017study}. 

The physics of locking and unlocking agreed on by various friction laws is, in short, 
either $\dot s$ equals $0$ or $T$ equals $\Phi$. 
Locking means rest ($\dot s=0$), while unlocking means the stress at the strength ($T = \Phi$). 
What the kinematic view of full coupling ($\dot s = 0$) and partial coupling ($\dot s > 0$) missed is the mechanics of unlocking $T = \Phi$, rather than the quiescence of locking $\dot s = 0$.

\subsection{Locking inversion as inverse crack problem}
\label{subsec:derivinglockinginversion}
The strength excess and slip rate are complementary on the frictional interface (eq.~\ref{eq:complementarity_strengthexcess}). In summary, it is the a priori definition of locking/unlocking as the pre-/post-yield phase. 
Equation~(\ref{eq:complementarity_strengthexcess}) itself depends on the behavior of $\Phi$, allowing for various estimates of locking in the inverse analysis. 
However, we can show below that, for quasi-stationary long periods (i.e., interseismic periods), the variety of those definitions vanishes, and they converge to a single formula (eq.~\ref{eq:complementarity_stressingsliprates}),  which sets the definition of interseismic plate locking uniquely.

The core of this claim is a one-paragraph proof. Specifically, we will show that the strength on the unlocked frictional interface is almost at the steady state when the assumption of quasi-stationarity (eq.~\ref{eq:quasistationary}) holds for a long period $t\in (0,\Delta t)$. Namely, `when $\Delta t\to\infty\cap \ddot s\simeq 0$, then $\dot \Phi\simeq 0\cup\dot s=0$'. 
The derivation is as follows.
When $T=\Phi$, then $\dot T=\dot \Phi$, 
so that when eq.~(\ref{eq:complementarity_strengthexcess}) holds, then
\begin{equation}
    (\dot \Phi-\dot T)\dot s= 0,
    \label{eq:complementarity_strengthexcessrate}
\end{equation}
which signifies the complementarity of the strength excess rate and the slip rate. 
Meanwhile, since the traction rate is proportional to the slip deficit rate (eq.~\ref{eq:slip2traction}), 
quasi-stationarity $\ddot s\simeq0$ (eq.~\ref{eq:quasistationary}) leads to $\ddot T\simeq 0$. 
When $\ddot T\simeq 0$ holds, eq.~(\ref{eq:complementarity_strengthexcessrate}) concludes $\ddot \Phi\simeq 0$ if $\dot s\neq0$, that is,
\begin{equation}
    \dot s\neq 0\Rightarrow \dot \Phi\simeq const.
    \label{eq:noninertialstrength}
\end{equation}
Now, the strength needs to satisfy eq.~(\ref{eq:noninertialstrength}) [$\Phi(t)\simeq \Phi(0)+\dot \Phi\Delta t$ if $\dot s\neq 0$], but the strength is positive and finite $\Phi\in(0,\Phi_{\rm max})$, where its upper bound $\Phi_{\rm max}$ is on the order of the normal stress, which is also positive and finite. For long periods, the limit of which is $\Delta t\to\infty$, such is possible only if
\begin{equation}
    \dot s\neq 0\Rightarrow \dot \Phi\simeq 0.
    \label{eq:steadystrength}
\end{equation}
That means, for quasi-stationary long periods, 
the strength is, on average, almost at a steady state when the interface is slipping.

Equations~(\ref{eq:complementarity_strengthexcessrate}) and (\ref{eq:steadystrength}) are followed by the complementarity of stressing and slip rates: 
\begin{equation}
    \dot T\dot s\simeq 0.
    \label{eq:complementarity_stressingsliprates}
\end{equation}
Thus, assuming a quasi-stationary interseismic period, eq.~(\ref{eq:complementarity_stressingsliprates}) was derived from eq.~(\ref{eq:yieldinglaw}) satisfied by many friction laws. 
Equation~(\ref{eq:complementarity_stressingsliprates}) is the same physical constraint of static-dynamic friction used in existing locking inversions. 
However, after this generalization, while $\dot s=0$ (locking) has the same meaning as that of static-dynamic friction, $\dot T=0$ (unlocking) is a condition expressing stationarity of strength rather than the manifestation of dynamic friction. 
This stationarity interpretation of $\dot T=0$ was introduced by \citet{funning2007asperities} as a hypothesis, and as above, this hypothesis is verified as a frictional behavior that does not depend on specific laws.

\begin{figure*}
   \includegraphics[width=120mm]{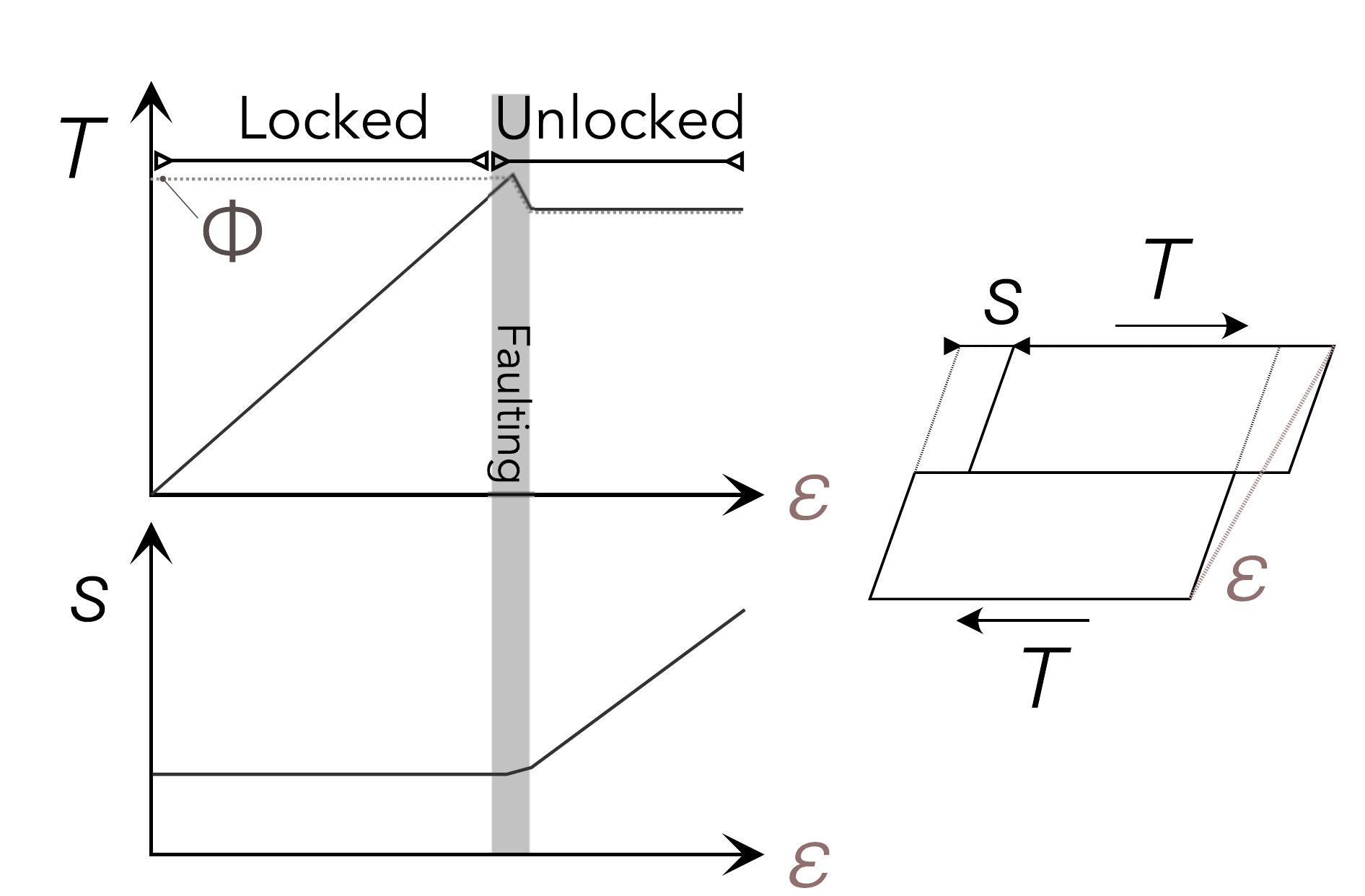}
  \caption{
Frictional behaviors under the yield criterion (eq.~\ref{eq:yieldinglaw}) and the complementarity between the rates of slip $s$ and stress loading $T$ (eq.~\ref{eq:complementarity_stressingsliprates}), exemplified by a biaxial test. 
Until the stress reaches its threshold $\Phi$, the stress increases in proportion to the strain $\epsilon$ without sliding (pre-yield: locked). 
After the stress reaches the strength, the interface slips so that the stress matches the strength (post-yield: unlocked). 
Different friction laws give different unlocked behaviors depending on the time evolution laws of the strength (eq.~\ref{eq:complementarity_strengthexcess}). 
Meanwhile, for the quasi-stationary behaviors outside the moment of faulting (gray in the figure), all friction laws give either zero slip rate or zero stress rate (eq.~\ref{eq:complementarity_stressingsliprates}), and the former is locked, and the latter is unlocked. 
}
  \label{fig:yielding_strength}
\end{figure*}

For intuitive illustration, suppose a biaxial test (Fig.~\ref{fig:yielding_strength}). 
The slip-stress curve of the crack face, which corresponds to the stress-strain curve of the bulk, is roughly divided into two phases: the locked phase, in which the stress responds to the strain increment in a Hookean manner, and the steady creeping phase, in which the strain increment is mostly compensated for by the slip of the crack face with fault stress unloaded. 
These correspond to $\dot s=0$ and $\dot T = 0$, the two phases of locking and steady unlocking (so to speak, stick and slip), respectively. 
The transient region between them (Fig.~\ref{fig:yielding_strength} gray) represents the unlocked phase outside the steady states. 
Many refinements of friction laws have been devoted to this transient, but negligible differences from the classical friction laws appear outside. 
That is what is meant by the fact that the slip rate and the stressing rate are complementary as per eq.~(\ref{eq:complementarity_stressingsliprates}) excluding that transient.
The crucial assumption of this model reduction is the long-term quasi-stationarity (eq.~\ref{eq:quasistationary} for a long period), often premised in coupling inversions. 

To summarize, in the most unified sense of friction, locking and unlocking are the terms to express the pre- and post-yield phases, respectively. 
Thus, as long as read in that sense, 
the locking/unlocking is an inherent status of frictional motions free from the assumptions of specific laws, allowing us to compare various forward and inverse models employing different friction laws. 
Moreover, interseismic locking is almost free from the differences in friction laws, 
so we can gauge the true-sense locking simply by using
the complementarity of slip and stressing rates (eq.~\ref{eq:complementarity_stressingsliprates}). 
The interseismic locking has almost no ambiguity both in its concept and measurement, having very small epistemic/model errors.

The forward model of the coupling inversion is the dislocation problem that specifies the slip on a crack face. 
In contrast, the forward model of locking inversion is the so-called crack problem that specifies the slip or stress in a mixed boundary condition. 
The stick-slip specification can be expressed by a binary, now called a locking parameter, denoted by $\Psi$.
The locking parameter is a Boolean expression of locking (1 is yes; 0 is no):
\begin{equation}
    \begin{aligned}
\Psi=1\Leftrightarrow \dot s=0 
\\
\Psi=0\Leftrightarrow \dot T=0 
    \end{aligned}
\label{eq:locking_constraint}
\end{equation}
$\Psi=1$ and $0$ represent locking (stick) and unlocking (slip), respectively. 
Locking inversions estimate the locking parameter $\Psi$ at each point on plate boundaries. 
Given that slip deficit inversions are sometimes also called locking inversions, one may refer to this locking inversion as stick-slip inversion. 
The observation equations of the locking inversion (the stick-slip inversion) consist of 
eq.~(\ref{eq:sdi_obseq}) on data and slip deficits, 
eq.~(\ref{eq:sfromsd}) on slip deficits and slips, 
eq.~(\ref{eq:slip2traction}) on slip deficits and stress, and eq.~(\ref{eq:locking_constraint}) on slips,  stress, and locking parameters. 
Equations~(\ref{eq:sfromsd}), (\ref{eq:slip2traction}), and (\ref{eq:locking_constraint})
express the slip field as a functional of the locking-parameter field. 
Then, the likelihood of the slip-deficit field given by eq.~(\ref{eq:sdi_obseq}) is converted to that of the locking-parameter field. 
This procedure becomes a simpler formula through fault subdivision, as summarized in \ref{sec:AppA}. 

Several applicability limits of the locking inversion should also be noted. 
As quantified in the above derivation, the interseismic phase is premised to be sufficiently long to exclude the non-quasi-steady unlocked zones (Fig.~\ref{fig:yielding_strength} gray).
In a precise sense, however, we cannot guarantee more than the smallness of the strength evolution rate, and thus $\dot \Phi\simeq 0$ does not mean the complete steady-state condition. 
The error of $\dot \Phi\simeq 0$ is $|\dot \Phi|$, which is bounded by the ratio of the strength upper bound to the interseismic period interval. 
Intuitively speaking, the nominal unlocked zones in locking inversions include the non-steady (but quasi-steady) unlocked zones, including the rim of unlocked zones surrounding the locked zones and very slowly accelerating nucleation zones. Those zones are not necessarily stable but rather unstable towards disruptive processes. 
$\dot T\simeq 0$ would mean stably creeping zones basically, and our discussions proceed basically under that recognition, but we must be aware of that proviso.
Another issue will be short-wavelength heterogeneity, which is neglected through the discretization, and short-time variations, which are neglected by the assumption of quasi-stationarity. 
As a first-order approximation, however, we now neglect those short-wavelength and high-frequency possibilities. 
One must also be aware that the binary view of locking and unlocking may be invalid outside friction (e.g., at a great depth), whereas the reduced concept of unlocking as force equilibrium keeps alive even in non-brittle rheology.

\subsection{Positional relationship of coupled, locked, and stressed zones}
\label{subsec:semantics}

Coupling, stressing, and locking are all indicators that represent different aspects of the fault state: slip, force, and friction. 
Since it was recognized that none of these indicators can substitute for the others, 
locking has been inferred using the working hypothesis of the Amontons-Coulomb friction. As we have pointed out, interseismic frictional behaviors can be well approximated by the Amontons-Coulomb friction, or precisely, by eq.~(\ref{eq:locking_constraint}) of the slip-rate-stressing-rate complementarity. 
Then, the solution of eq.~(\ref{eq:locking_constraint}) will help to interpret these three indicators in relation to each other. 

\begin{figure*}
    \centering
   \includegraphics[width=130mm]{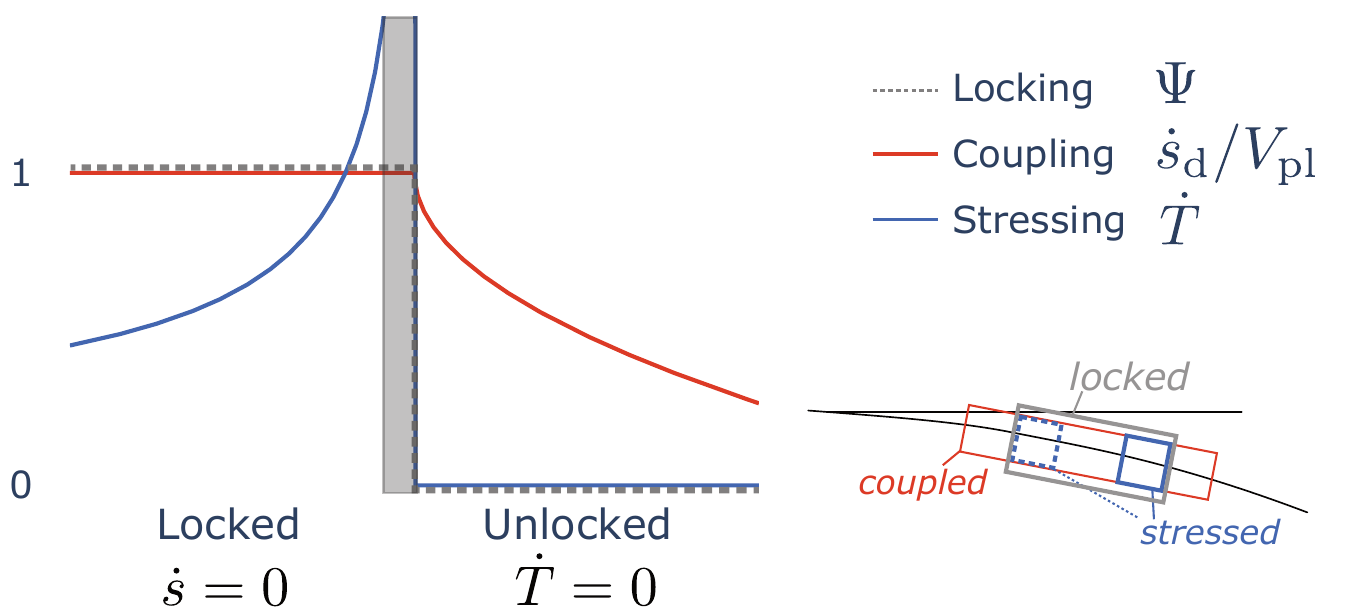}
  \caption{
  Spatial patterns of coupling, locking, and stressing, expected from the slip-rate-stressing-rate complementarity (eq.~\ref{eq:locking_constraint}). 
  A typical two-dimensional solution is visualized with a schematic, especially around the boundary of a locked zone and an unlocked zone. 
  The gray region masked in Fig.~\ref{fig:lcs_rel} corresponds to the gray region in Fig.~\ref{fig:yielding_strength} and represents the very vicinity of the locked zone tip, to which eq.~(\ref{eq:locking_constraint}) does not apply due to the artifact of divergent stress above the strength.
  } 
  \label{fig:lcs_rel}
\end{figure*}

Figure~\ref{fig:lcs_rel} describes spatial patterns of coupling, locking, and stressing on a frictional surface governed by eq.~(\ref{eq:locking_constraint}). 
A planar fault in a homogeneous isotropic two-dimensional full space is considered. 
Here, coupling corresponds to conventional kinematic coupling, 
locking corresponds to the mechanical coupling in the sense of \citet{herman2018accumulation}, 
and stressing corresponds to the mechanical coupling in the sense of \citet{saito2022mechanically}.

Equation~(\ref{eq:locking_constraint}) imposes zero slip deficit rates inside the asperity while imposing zero stress rates outside. 
This boundary condition is parallel to the standard crack problem that imposes 
zero slip (and thus zero slip gradient) outside the asperity
while imposing zero stress inside. 
This similarity of the boundary condition on the dislocations (slip gradients) and stress results in 
similar solutions of eq.~(\ref{eq:locking_constraint}) and orthodox crack problems (Fig.~\ref{fig:lcs_rel}). 
On a planar two-dimensional fault, 
the Hilbert transform, which denotes the convolution of a given function and the signed inverse distance divided by $\pi$, converts the dislocation to the traction normalized by the effective stiffness~\citep[e.g.,][]{rubin2005earthquake}. 
Thus, zooming in on the boundary of locking and unlocking, the associated solutions for both the dislocation and normalized traction become the real part of the inverse square root distance from the locking-unlocking boundary, which is converted to its sign-flipped mirror image through the Hilbert transform. 
The proportionality constant of this solution is determined by the condition outside the crack tip, occasionally remarkably reduced just beneath the trench~\citep{herman2018accumulation}.

This solution of eq.~(\ref{eq:locking_constraint}) indicates 
a positional relationship of the coupled, stressed, and locked zones (Fig.~\ref{fig:lcs_rel}).
The coupling is one inside the locked zone and gradually decreases outside the locked zone, roughly inversely proportional to the square root of the distance from the locked zone tip. 
The stress concentrates around the locked zone tip, 
and the stressed zone is localized inside the locked zone. 
Briefly, the locked zone concentrates the stress around the tip, deforms the matrix surrounding the tip, and slides the proximate unlocked zone. 
Consequently, the boundary of the locked zone and the unlocked zone is located at the intersection of a highly coupled zone and a highly stressed zone. 
Since eq.~(\ref{eq:locking_constraint}) is based on a very robust equality of eq.~(\ref{eq:complementarity_stressingsliprates}) as shown in the previous subsection, this positional relationship is universally expected to interseismic frictional sliding. 
More precisely speaking, as expressed by the non-negativity of coupling and stressing in \citet{lindsey2021slip}, when the slip directions are invariant, 
stressing is positive inside the locked zone and zero outside, while
coupling is full inside the locked zone and positive outside.

Furthermore, conventional coupling inversions impose the smoothing prior of slip deficits, while the stressing inversion imposes traction damping prior~\citep{saito2022mechanically}. 
Therefore, when comparing the results of coupling inversions and stressing inversions using different prior constraints, the estimated coupled zone tends to widen and the estimated stressed zone tends to narrow, conceivably emphasizing this positional relationship of coupled, locked, and stressed zones, as confirmed in our benchmark analysis (\ref{subsec:appresult}). 
Obviously, the influence of prior constraints is not that simple always. \citet{lindsey2021slip} show that a prior constraint on the stressing, that of non-negativity in their case, can capture coupled zones undetected when using smoothing constraints on coupling, demonstrating that the constraint on stress loading can also widen the estimated coupled zone. More fundamentally, \citet{herman2018accumulation} and \citet{lindsey2021slip} demonstrate that beneath-trench/trough unlocked zones may be misinterpreted as shallow extensions of the locked zones in the presence of the stress shadows of the asperity. The shallow portion of a locked zone may be undiscussable by observed data alone, that is, essentially within the realm of the prior constraint, although discussing it is far beyond the scope of this paper.

As above, stress concentration around locked-zone tips and the resultant positional relationship of coupling, locking, and stressing are widely expected in frictional sliding, but the stress divergence right at those tips (e.g., cohesive zones) is the artifact of eq.~(\ref{eq:locking_constraint}) 
because the yield criterion expects the stress below the strength (eq.~\ref{eq:yieldinglaw}). If eq.~(\ref{eq:yieldinglaw}) is read in the Amontons-Coulomb sense, it violates the original criterion itself. 
This artifact produces higher stress for finer meshes in locking inversions. 
Meanwhile, even with this divergent solution, the strain energy density is finite~\citep{freund1998dynamic}. 
Therefore, whereas eq.~(\ref{eq:locking_constraint})-based inference of locking inversions fails to evaluate the stressing rate in the very proximity of the crack tips, it can evaluate the strain energy release rate even within those apparently stress-divergent zones. 
Microscopic details of crack tips have been treated in that manner in classical fracture mechanics~\citep{rice1968path}. 

This artificial stress divergence makes the solution of eq.~(\ref{eq:locking_constraint}) inaccurate in post-yield transient (unlocked but non-steady) zones, the widths of which depend on the fault properties. 
Interseismically, those zones would correspond to $a\sim b$  (more accurately, conditionally stable) in the RSF, 
and some physics-based models suggest the seismogenic zones of slow earthquakes may be 
$a\sim b$ areas with finite width~\citep[e.g.,][]{liu2007spontaneous}. 
As such a hypothesis is outside the applicability of the locking inversion, 
\citet{bruhat2017deformation} include the post-yield transient zone in their model, although the transient zone physics in their model is the asperity erosion, quasistatic propagation of an unlocking front, rather than cohesive forces making crack tip stress finite. 
Geodetic inversions by \citet{sherrill2024locating} using the \citet{bruhat2017deformation} model showed that 
the width of that transient zone depends on the tectonic setting. 
According to their results, neglecting post-yield transient zones (eq.~\ref{eq:complementarity_stressingsliprates}) is a good approximation for the current state of the Nankai subduction zone we later investigate, as revisited in the discussion section.

\section{A transdimensional scheme of locking inversion}
\label{sec:transdim}

The epistemic errors, also called model error/bias, were found to be small enough in locking inversions for quasi-stationary interseismic phases. However, it does not negate that the inverse problem of locking is challenging. 
As in many distributed slip inversions, 
likelihood-based approaches of locking-parameter fields easily overfit to data~\citep{herman2020locating}. 
The use of prior information is one way to avoid this issue, but adjusting the prior constraint is harder than in coupling inversions~\citep{johnson2010new}. 
Additional computational difficulties also arise in locking inversions due to the nature of discrete optimization problems in mathematics, to which the locking inversions belong. 
For tractable estimations of locking, we now construct a transdimensional scheme \citep{dettmer2014trans} of locking inversions, which varies the number of basis functions. Basis functions here imitate locked segments, termed the asperities in fault mechanics~\citep{barbot2019slow}. 

\begin{figure*}
   \includegraphics[width=140mm]{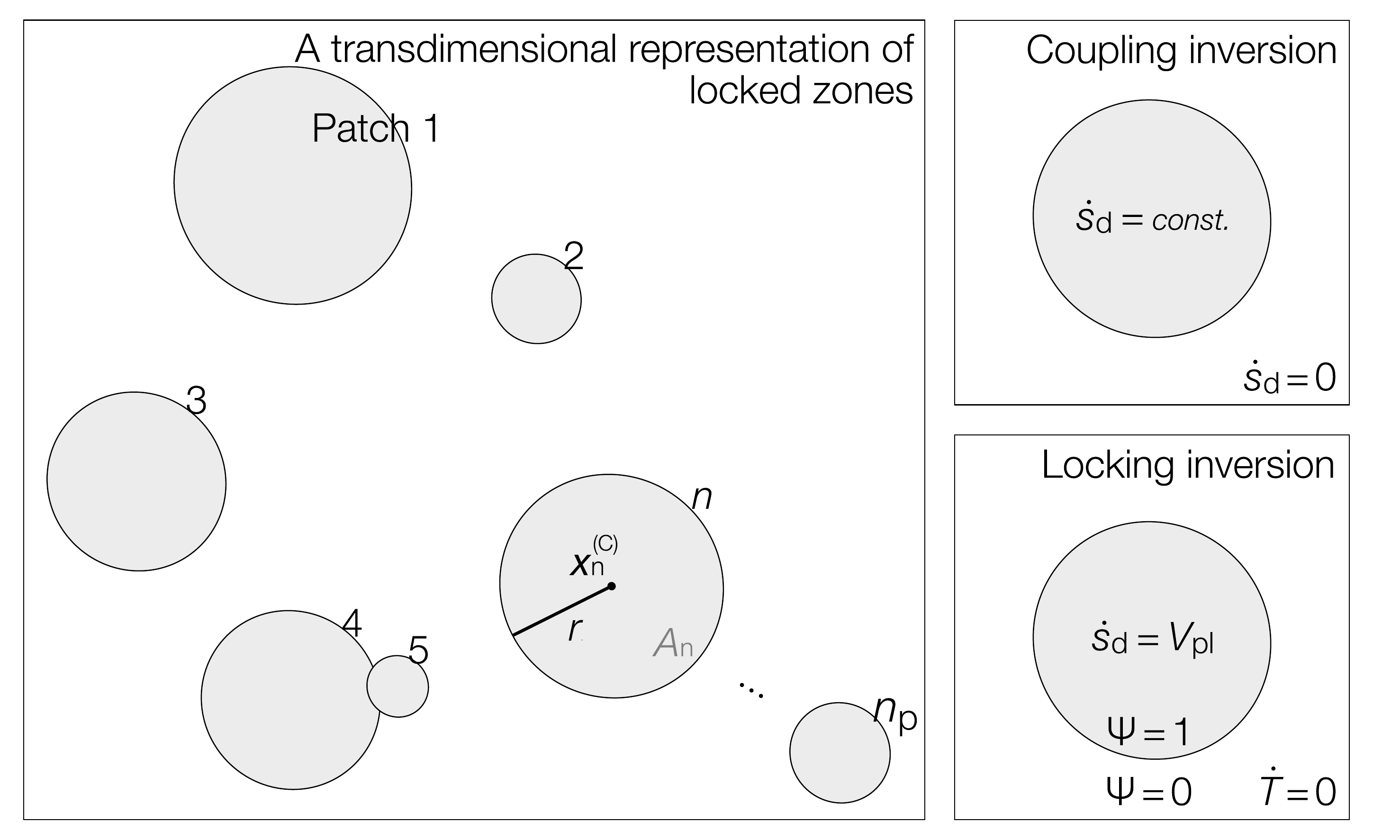}
\caption{
Transdimensional parametrization of locked zones. 
Locked zones are decomposed into 
$n_{\rm p}$ segments, denoted by $A_n$, 
parametrized by center locations ${\bf x}_n^{\rm (C)}$ and radii $r_n$. 
While transdimensional schemes of slip deficit inversions (here called coupling inversion) 
superpose constant slip-deficit-rate zones on a slip-deficit-free boundary, transdimensional locking inversions superpose locked segments, where $\dot s_{\rm d}=V_{\rm pl}$, on a traction-free boundary, where $\dot T=0$. 
}
\label{fig:parametrizinginterlockingpattern}
\end{figure*}

Figure~\ref{fig:parametrizinginterlockingpattern} is the method schematic. 
The locked zone is decomposed into segments $A_n$, within which the fault is locked ($\Psi=1$):
\begin{equation}
\Psi(\boldsymbol\xi)= 
\begin{cases}
1  & \boldsymbol\xi\in \sum_n A_n
\\
0 & {\rm otherwise}
\end{cases}
\label{eq:transdimrepofpsi}
\end{equation}
By employing those segments as basis functions, 
our scheme maps the spatial-pattern estimation of the locked zone to the configuration estimation of frictionally locked segments. 
We parametrize those segments by circles, in the spirit of \citet{kikuchi1982inversion} for simplicity, although transdimensional schemes often utilize Voronoi cells~\citep{dettmer2014trans,tomita2021development}. 
The center locations $ \boldsymbol\xi_n$ and radii $r_n$ of asperities, denoted by $\{ \boldsymbol\xi_n,r_n\}_{n=1,...,n_{\rm p}}$, are the model parameters of this scheme. 
The number of asperities $n_{\rm p}$ works as an additional parameter to specify the structure of the model, namely the hyperparameter of this scheme. 
Note that altering the numbering rule of asperities (e.g., shuffling their numbers) does not affect the locked zone pattern. Therefore, to estimate the locking-parameter field from the configuration of asperities, we must consider the permutation of asperities, not their combination. In this study, we have implemented it by employing a sorting of asperities, or specifically, by sorting them according to the lateral position.

The above transdimensional locking inversion is based on the superposition of simple solutions as in transdimensional coupling inversions 
(Fig.~\ref{fig:parametrizinginterlockingpattern}). 
Reducing the degrees of freedom results in 
discarding the inversion resolution. This approach can extract robust information in return for discarding error-prone details.

Once finishing the above transdimensional parametrization of the locking-parameter field (eq.~\ref{eq:transdimrepofpsi}, with $\boldsymbol\xi\in  A_n \Leftrightarrow | \boldsymbol\xi- \boldsymbol\xi_n| < r_n$ assumed), 
the remaining is the same as the conventional grid-base locking inversions. 
We assume elementwise-constant subdivision of $\dot s_{\rm d}$, $ \dot s$, and $\Psi$, with the center collocation of $\dot T$. 
Then, the observation equation of slip deficits (\ref{eq:sdi_obseq}) is discretized as follows: 
\begin{equation}
    {\bf d}={\bf H}\dot {\bf s}_{\rm d}+{\bf e},
    \label{eq:discreteobseq_slipdeficit}
\end{equation}
where ${\bf d}$ and ${\bf e}$ are vector notations of $\bar d_i$ and $e_i$, respectively, ${\bf H}$ represents the  discrete form of Green's function $G$, and $\dot {\bf s}_{\rm d}$ denotes the slip deficit rates of fault elements. 
The probability of the error term (eq.~\ref{eq:GaussianError}) sets the likelihood $L(\dot {\bf s}_{\rm d})$ of $\dot {\bf s}_{\rm d}$: 
\begin{equation}
    L(\dot {\bf s}_{\rm d})=\mathcal N({\bf H}\dot {\bf s}_{\rm d},{\bf C}_{\bf e}).
\end{equation}
Hereafter, $L(\cdot):=P({\bf d}|\cdot)$ denotes the likelihood.
Next, we relate the slip deficit rates of elements to the locking parameters of elements. 
Equation~(\ref{eq:sdpsi}) in \ref{sec:AppA} is a discrete expression of the slip-deficit-rate field $\dot {\bf s}_{\rm d}(\boldsymbol\Psi)$ given the discretized locking-parameter field $\boldsymbol\Psi$, where $\boldsymbol\Psi$ denotes a vector storing the locking-parameter values of elements. 
Substituting ${\bf s}_{\rm d}={\bf s}_{\rm d}(\boldsymbol\Psi)$ into $L(\dot {\bf s}_{\rm d})$, 
we obtain the likelihood of the discrete locking-parameter field: 
\begin{equation}
L(\boldsymbol\Psi)= \mathcal N({\bf H} \dot {\bf s}_{\rm d}(\boldsymbol\Psi),{\bf C}_{\bf e}).
\end{equation}
Note that $\dot {\bf s}_{\rm d}$ and $\boldsymbol\Psi$ have a one-to-one correspondence, given the uniqueness of solution in crack problems. 

The locking parameters $\boldsymbol\Psi$ are now given by a function $\boldsymbol\Psi(\{ \boldsymbol\xi_n,r_n\}_{n=1,...,n_{\rm p}})$ of the asperity configuration $\{ \boldsymbol\xi_n,r_n\}_{n=1,...,n_{\rm p}}$. 
The locking-parameter field and the asperity configuration do not have one-to-one correspondence because small asperities buried beneath large asperities do not affect the locking-parameter field. 
To avoid a problem complicated, we introduce an additional rule that asperity configurations are identified if the $\Psi$ field is the same so that one-to-one correspondence between the asperity configuration ($\{ \boldsymbol\xi_n,r_n\}_{n=1,...,n_{\rm p}}; n_{\rm p}$) and the locking-parameter field ($\boldsymbol\Psi$) holds: 
\begin{equation}
    L(\{ \boldsymbol\xi_n,r_n\}_{n=1,...,n_{\rm p}}; n_{\rm p})\approx L(\boldsymbol\Psi). 
    \label{eq:likelihood_asperity_configuration}
\end{equation}
Equation~(\ref{eq:likelihood_asperity_configuration}) is a conversion formula to transform the likelihoods of different series expansions of locking-parameter fields, 
since asperity configuration is one of the series expansions of a locking-parameter field. 
Then, for brevity, 
the left-hand side $L(\{ \boldsymbol\xi_n,r_n\}_{n=1,...,n_{\rm p}}; n_{\rm p})$ of eq.~(\ref{eq:likelihood_asperity_configuration}) may also be denoted by $L(\Psi; n_{\rm p})$.

To solve the above transdimensional problem, 
we now conduct an objective point estimation of the hyperparameter $n_{\rm p}$. 
Bayesian information criterion~\citep[BIC;][]{schwarz1978estimating} states that 
the marginal likelihood of the hyperparameter $n_{\rm p}$ is 
given by the conditional maximum likelihood of the model parameters minus the penalty term proportional to the number of model parameters (now 3$n_{\rm p}$), weighted by 
the log number of data divided by 2: 
\begin{equation}
\ln L( n_{\rm p})\simeq
\mbox{max}_{\Psi}
\ln L(\Psi; n_{\rm p})- \frac{3\ln N}2 n_{\rm p}
\label{eq:BICapprox}
\end{equation}
The optimization function of the BIC is $-2\ln L( n_{\rm p})$ evaluated by eq.~(\ref{eq:BICapprox}). 
Equation~(\ref{eq:BICapprox}) is Laplace's approximation using a Gaussian approximation of the distribution around its peak, 
thus being a rough approximation for multimodal distributions. The likelihood of locking parameters (asperity configurations) is actually multimodal (\S\ref{subsubsec:lockinginv3}). 
Nonetheless, similar Laplace's approximation is adopted in practice and works well to some extent for multimodal distributions, such as in the epidemic-type aftershock-sequence model in statistical seismology~\citep{ogata1990monte}, and thus we rely on eq.~(\ref{eq:BICapprox}).

In summary, our scheme is the maximum-likelihood estimation of the asperity configuration for a given number of asperities $n_{\rm p}$, and $n_{\rm p}$ is optimized by the BIC. In solving this likelihood maximization, we have employed a few numerical tricks, which are summarized in \ref{sec:AppC}. 
See the supplement described in the Open Research Section for code snippets. The key to this scheme is mapping a discrete optimization (locking of each element) to a continuous optimization (asperity configuration), which allows us to use common optimization methods for continuous variables. Similar courses of implementation can be found in the use of belt-shaped locked zones (i.e., a long polygonal asperity) in \citet{kimura2021mechanical} and \citet{sherrill2024locating}. Extensions to Voronoi cells and mechanically favorable ellipses are also conceivable. Since the scope of our method development is in the first-order model, however, we limit our considerations to circular asperity.

\section{Application}
\label{sec:Application}

Megathrust earthquakes recur along the Nankai subduction zone off southwestern Japan, driven by the northeastward subduction of the Philippine Sea Plate beneath the Amur Plate~\citep{ando1975source}. 
Paleoseismic records suggest several sections of Mw 8 class asperities aligned along this subduction zone~\citep{ishibashi2004status,furumura2011revised,garrett2016systematic}.
Such hypothetical asperities and their partitions are supported by velocity anomalies~\citep{kodaira2000subducted,kodaira2006cause}, gravity anomalies~\citep{wells2003basin}, and 
estimated slip zones of megathrust earthquakes~\citep{kikuchi1982inversion,murotani2015rupture}. 
Recent findings of slow earthquakes allow us to assess this hypothesis of aligned asperities in light of seismic activities~\citep{obara2016connecting}. 
Modern approaches of geodetic inversions enable the direct inference of locked zones~\citep{kimura2021mechanical,sherrill2024locating}. Here, we revisit the locking inversion of the Nankai to characterize the locked zone as asperities. 
The primary aim of the following analysis is to verify the developed scheme, specifically whether it extracts robust long-wavelength properties.

\subsection{Data}
\label{subsec:data}

We invert the averaged horizontal-velocity data of the onshore Global Navigation Satellite System (GNSS) and offshore Acoustic GNSS (GNSS-A) processed by \citet{yokota2016seafloor} (Figs.~\ref{fig:bestlockingcoupling} and \ref{fig:kinematiccouplinginversion}). 
The data period for onshore GNSS is from March 2006 to December 2009, which is a snapshot of the interseismic period of the Nankai subduction zone during which only a small number of large $M \geq 6$ earthquakes occurred. 
The data period for offshore GNSS-A is from 2006 to 2016, and GNSS-A data are fitted by M-estimation regression with postseismic deformation of the 2011 Mw 9.0 Tohoku-Oki earthquake removed. 
The number of observation points is 261, and we use two horizontal components. The number of data $N$ is 522. 

According to coupling inversions, both for synthetic and actual data, 
slip deficits of the shallow boundary 20--30 km away from the GNSS-A stations are left unresolved~\citep[Extended Data Figure 6][also, see Fig.~\ref{fig:kinematiccouplinginversion}]{yokota2016seafloor}. It corresponds to the areas within a few fault elements near the trough axis; the subdivided fault geometry appears later in Fig.~\ref{fig:bestlockingcoupling}. 
The slip-deficit patterns on those shallow elements are largely extrapolated by the prior constraints or basis functions.

\subsection{Problem setting}
\label{subsec:probset}

During the evaluation of slip deficit and stress (eqs.~\ref{eq:sdi_obseq} and \ref{eq:slip2traction}) in our locking inversion, the medium is approximated by a half-space homogeneous isotropic Poisson solid with the fault geometry of the Japan Integrated Velocity Structure Model version 1~\citep{koketsu2009proposal,koketsu2012japan}. The approximated ground surface of the half space is set at sea level. 
In the half-space model, we assume a stiffness of 40 GPa when computing stressing rates, supposing typical shear wave speeds around 3.5 km/sec and mass densities around 3 g/cm${}^3$, although coupling and locking inversions do not require specific values of stiffness. For simplicity, slip deficits across the plate boundary are approximated to be parallel to the nominal subduction direction of N55$^\circ$W. 
The value of $V_{\rm pl}$ here refers to the plate model of MORVEL2010~\citep{demets2010geologically}, acquired from that plate model along the trough axis and extrapolated along subduction, which is around 7 cm/yr.

The assumption of half-space homogeneous elasticity \remove{is obviously inaccurate}\add{could produce model bias in the surface displacement and the fault stress}, so we have checked Green's function errors (\ref{subsec:appresult}). In the benchmark test of \ref{subsec:appresult}, we have employed the displacement Green's function developed and distributed by \citet{hori2021high}, which is based on the Japan Integrated Velocity Structure Model version 1, accounting for topography, elastic heterogeneity, and the roundness of the earth. We confirmed that the half-space and high-fidelity models led to similar results in coupling inversions, although differences exist in the absolute values of coupling as well as in the coupling pattern at the eastern edge of the Nankai subduction zone. 
Our locking inversion is limited to half-space analysis as the traction Green's function is not included in \citet{hori2021high}. 
The resolution tests have been conducted by \citet{yokota2016seafloor} for the coupling inversion of the same data for the same study area~\citep[Extended Data Figure 6 in][]{yokota2016seafloor}, and we do not repeat them.

We account for the off-fault physics unmodeled in the assumed Green's function as Green's function errors. We follow the method of \citet{yagi2011introduction} and treat the error term ${\bf e}$ as a summation of observation errors and Green's function errors, both of which are approximated by Gaussian variables independent of each other, now expressed as 
\begin{equation}
    {\bf e}\sim \mathcal N({\bf 0},\sigma^2 {\bf I}+\Sigma^2 {\bf Hs}_{\rm d}{\bf s}_{\rm d}^{\rm T}{\bf H}^{\rm T}),
    \label{eq:errorterm_explicit}
\end{equation}
where $\sigma^2$ and $\Sigma^2$ are scale factors that represent the magnitudes of observation errors and Green's function errors, respectively. 
Because data include Green's function errors multiplied by slip (deficits), the error term ${\bf e}$ depends on the model parameters. 
The proportionality between Green's function errors and Green's function expresses the fact that path effects and site effects are generally proportional to Green's function itself~\citep{yagi2011introduction}; the primary cause of Green's function errors for our analysis is suggested to be off-fault inelastic deformation from coupling inversions (\ref{subsec:appresult}). 
The estimation method of data covariance in eq.~(\ref{eq:errorterm_explicit}) is established in slip inversions~\citep{yagi2011introduction} based on Akaike's Bayesian information criterion~\citep[ABIC, here the same role as model likelihood;][]{akaike1980use,yabuki1992geodetic}. In this study, we first estimate the data covariance ($\sigma^2 {\bf I}+\Sigma^2 {\bf Hs}_{\rm d}{\bf s}_{\rm d}^{\rm T}{\bf H}^{\rm T}$ in eq.~\ref{eq:errorterm_explicit}) from the optimal coupling inversion (our benchmark estimate Fig.~\ref{fig:kinematiccouplinginversion}a) according to \citet{yagi2011introduction} and use this data-covariance estimate when performing the locking inversion.
As developed in \S\ref{sec:transdim},
the optimal solution of our locking inversion is the maximum-likelihood estimation given the number of asperities selected by the BIC. 

To compare the half-space model and high-fidelity model mentioned above,
the fault in our half-space model is triangulated with 20 km intervals, which is 
the knot interval of the B-spline basis functions for slip distributions in \citet{hori2021high}. 
This mesh/knot interval is roughly equal to or smaller than the intervals of the offshore GNSS-A data we used. 
The finer spatial scale is outside the scope of this study. 
Mesh removals of overly obtuse triangles are made for numerical stability, which resulted in missing meshes in a very shallow area within 20 km of the trough axis; this could be problematic but is irrelevant in the following analysis because those zones are outside the data coverage. 
Accordingly, the asperity radii are assumed to be horizontally larger than 20 km in the following application.
The asperity radii are measured in horizontal dimensions for simplicity.  
The intersections of circles and the plate boundary are identified as the locked zone. See \ref{sec:AppC} for computational details.

An essential model limitation comes from the complexity of subduction. In the Nankai subduction zone, the Amur Plate collides with the North American Plate, and the Izu Microplate moves relative to the Philippine Sea Plate, affecting the recurrent-interval evaluation of the megathrust earthquakes in the Nankai subduction zone ~\citep{heki2001plate,miyazaki2001crustal}. 
Such unmodeled but significant long-wavelength perturbations may change the results~\citep{loveless2010geodetic}, but precisely considering them is future work for this study.

\subsection{The optimal estimate of plate locking}
\label{subsubsec:lockinginv1}

\begin{figure*}
   \includegraphics[width=140mm]{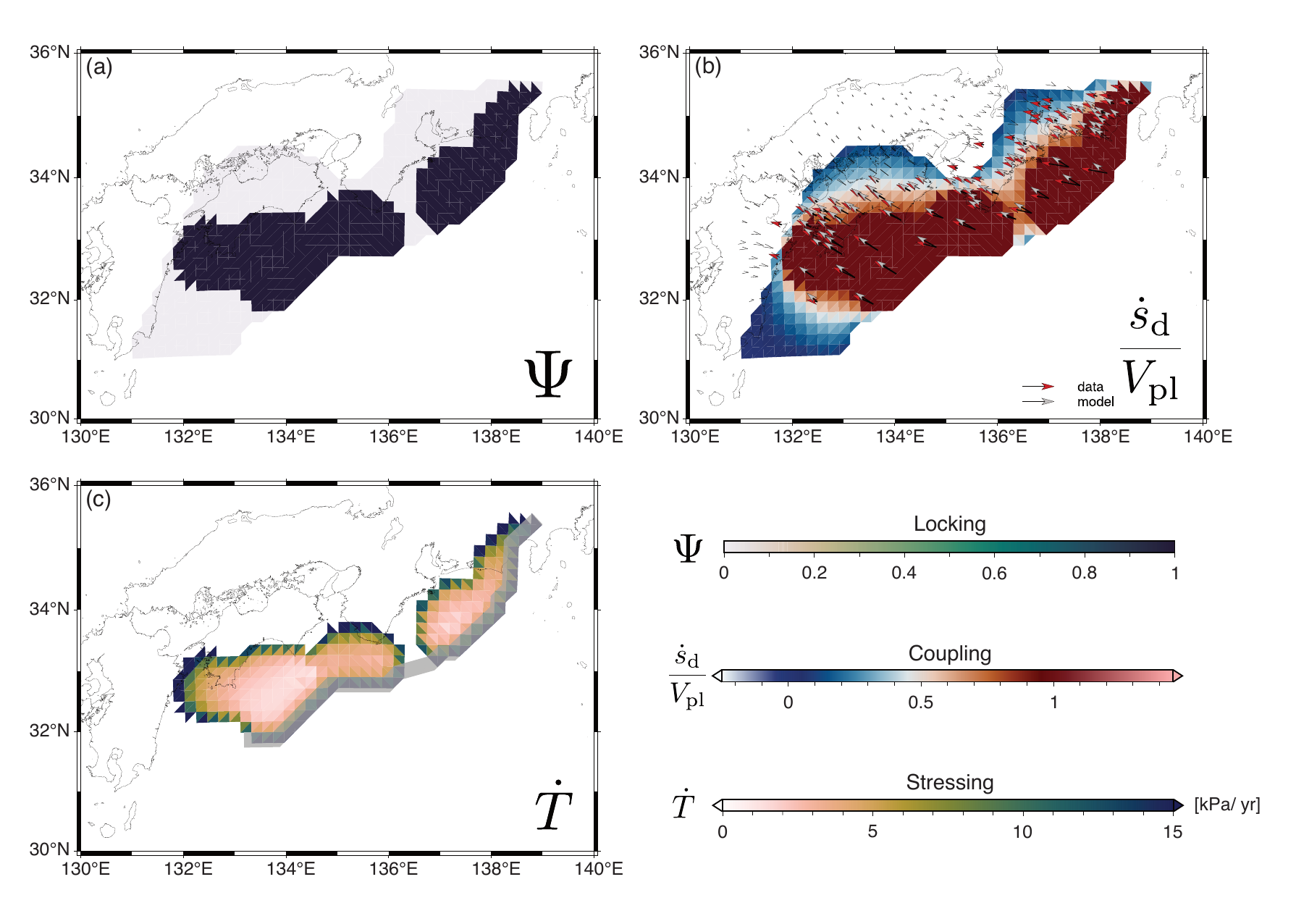}
  \caption{
  The optimal estimate of our locking inversion. Maximum likelihood estimation is employed for the configuration optimization of circular asperities. The number of asperities is optimized by the BIC.
  (a) The optimal locking-parameter field $\Psi$. 
  (b) The coupling field $\dot s_{\rm d}/V_{\rm pl}$ computed from the optimal locking-parameter field. Observed and modeled surface displacements are indicated by arrows.
  (c) The stressing field $\dot T$ computed from the optimal locking-parameter field. 
  Stressing rates greater than 15 kPa/year are rounded, given the unresolvable stress concentration at the crack tip. Artificial stress concentration right beneath the trough is masked for visibility.
}
  \label{fig:bestlockingcoupling}
\end{figure*}

The optimal solution of our locking inversion is summarized in Fig.~\ref{fig:bestlockingcoupling}. 
Locking inversion depicts how locking, coupling, and stressing, three similar concepts, are spatially allocated. Full coupling, that is, locking of the seismogenic depth contrasts with full creeping of great depths that drives the earthquakes recurrently (Fig.~\ref{fig:bestlockingcoupling}b). The mid-depth is coupled even if unlocked because of the pinning effects from the locked zones distributed over the seismogenic depth (Fig.~\ref{fig:bestlockingcoupling}a, b). Stress loads are localized within the locked zone, and a significant stress concentration occurs around their edges (Fig.~\ref{fig:bestlockingcoupling}c). 
Stress loads near the surface are effectively canceled because of the half-space effect, although it is unmodeled and thus masked in our analysis, where the fault does not intersect the ground and always produces artificial stress concentration right beneath the trough (Fig.~\ref{fig:bestlockingcoupling}c, masked).

The causality of kinematic and mechanical couplings is also clear in the locking inversion. 
In locking inversion, the model parameter represents the locking-parameter field (Fig.~\ref{fig:bestlockingcoupling}a). The estimate then sets the boundary condition of constant slip or stress and determines the slip deficit (Fig.~\ref{fig:bestlockingcoupling}b) and stress loading (Fig.~\ref{fig:bestlockingcoupling}c).
The locking inversion shows that locking is the cause and slip deficit as coupling and stress loading as stressing are the effects.

The optimal slip-deficit (coupling) field (Fig.~\ref{fig:bestlockingcoupling}b) 
reproduces major features of our benchmark solution (Fig.~\ref{fig:kinematiccouplinginversion}a) constructed from kinematic coupling inversions: 
the western and eastern subdomains of full coupling and high coupling at depth around the Bungo Channel. 
Five asperities are estimated. 
From the west, 
(i) Bungo-Channel plus Hyuga, (ii \& iii) Nankai, 
(iv) Tonankai, and (v) Tokai.

Those asperities constitute a belt of locking within the seismogenic depth, but this belt is separated into the eastern and western parts. 
This gap is blurred in coupling and clearer in stressing.
Since the locking gap is a source quadrupole arising from a neighboring dipole pair (i.e., adjacent locking-unlocking boundaries), the stress concentration and the high coupling coexist around the locking gap.

An advantage of locking inversions over coupling inversions is seen the reproducibility of stress concentration hard to resolve in kinematic approaches, which is the same as the detectability of the locking boundary. 
For visibility, we round the divergent stressing rates at the crack tips (blue areas in Fig.~\ref{fig:bestlockingcoupling}c) to 15 kPa/year, which is roughly 
half an order of magnitude larger than the 3--6 kPa/year stressing rate achieved around the center of the asperities. 
Even without rounding, the stressing rate at the crack tip is necessarily an approximate value (\S\ref{subsec:semantics}).
We can recognize stress loading around 3--6 kPa/year inside the stressed patches.
Higher stress loading near the locked zone tips (i.e., stress concentration) is a physically expected feature hard to grasp in kinematic inversions. 

Recalling that the highest stressing rate at the crack tip (blue areas in Fig.~\ref{fig:bestlockingcoupling}c) is determined by subdivision lengths in locking inversions (\S\ref{subsec:semantics}), 
even our locking inversion could truncate shorter-wavelength natures within each element, and thus the stress concentration will be more intense in reality.
The cohesive zone width is thought to be at most on the order of kilometers, which corresponds to the critical slip-weakening distance around centimeters, thus being a one-digit times smaller than our mesh size, although discussions remain in terms of slow earthquake source physics, as referred to in the next section.

Regarding the areas right beneath the trough, 
our half-space model sets the virtual ground surface at the sea level well above the trough axis, inducing an effective constraint of zero coupling just beneath the virtual ground surface. 
Some portions of the beneath-trough stress concentration may be true, inducing shallow slow earthquake activity, but
our model setting is unable to discuss them. Still, that artificial stress concentration right beneath the trough is now distant from the unlocked zone, not affecting the slip deficit pattern, thus irrelevant to the current data fitting.

A trivial difference between locking inversions and coupling inversions exists in the definition range of coupling, which is typically $(-\infty,\infty)$ in conventional coupling inversions and $(0,1]$ in locking inversions (Fig.~\ref{fig:bestlockingcoupling}b). 
Coupling inversions estimate slip deficits as model parameters, and observation errors and Green's function errors may drive the estimated coupling out of $(0,1]$. 
Locking inversions, where the model parameters denote the locking-parameter field, set the coupling field from the locking-parameter field via eq.~(\ref{eq:complementarity_stressingsliprates}), and the coupling ratios in locking inversions are forced to be within $(0,1]$. 
Since this difference comes from a priori constraints, and forcing the coupling within $(0,1]$ is sometimes employed in the coupling inversions,
we do not delve into the coupling ratios outside $(0,1]$. 
\add{A negative coupling and an above-one coupling are often interpreted as full sliding and full coupling, respectively, so comparisons between locking and coupling inversions would be acceptable.}

\subsection{Statistical behaviors of estimates}
Our optimal estimate of locking inversion (Fig.~\ref{fig:bestlockingcoupling}) is supported by our benchmark solution of coupling inversion (\S\ref{subsubsec:lockinginv1}) while it also reveals the locking gap, which is blurred in a pure kinematic view of coupling inversion. Additional verification then becomes necessary, as shown below. 

\subsubsection{Likelihood landscapes}
\label{subsubsec:lockinginv2}

\begin{figure*}
    \includegraphics[width=120mm]{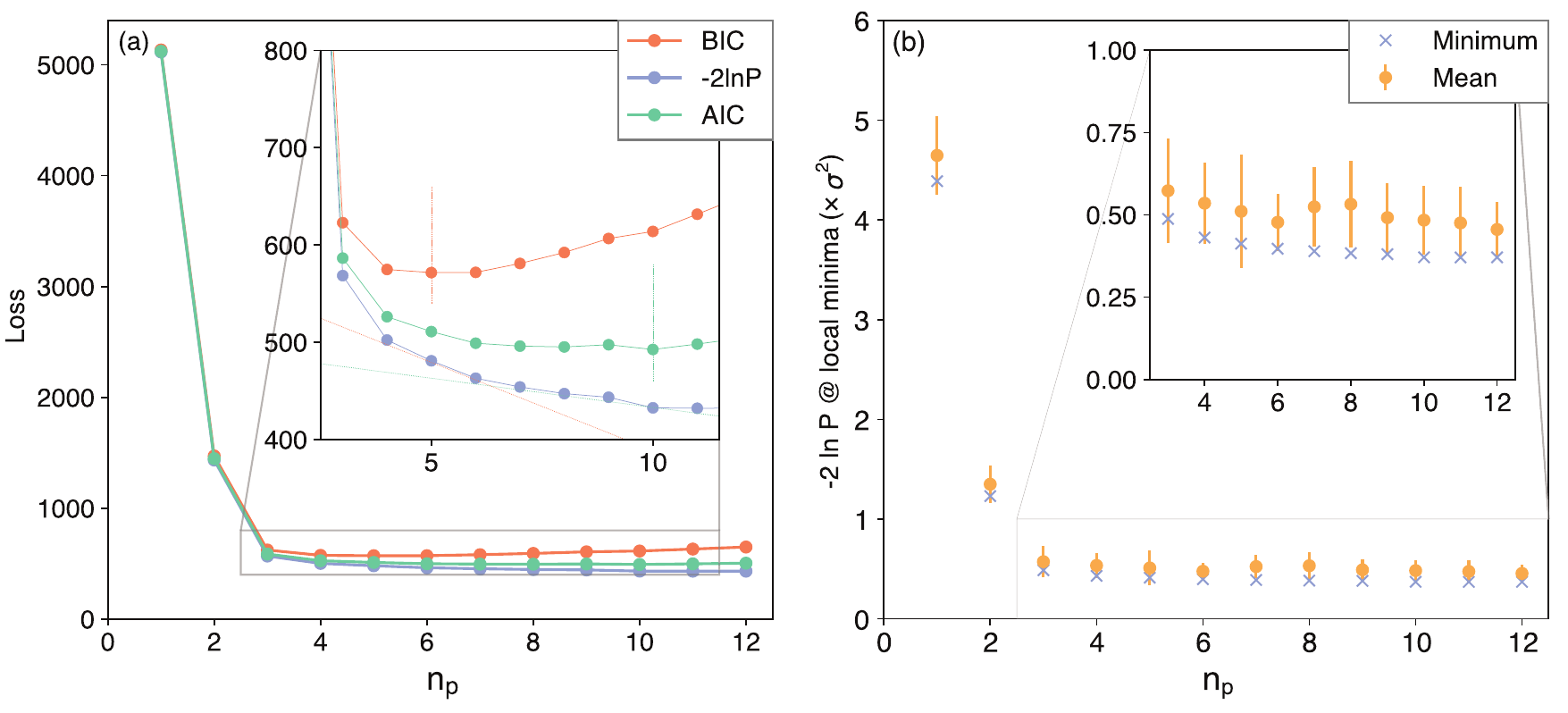}
  \caption{
  Probability landscapes of locking inversions.
  (a) The likelihood $L(n_{\rm p})$ of the number of asperities $n_{\rm p}$, approximately evaluated by the BIC (orange). 
  The conditional maximum log likelihood $\max_\Psi \ln L(\Psi;n_{\rm p})$ given $n_{\rm p}$ (blue) and AIC (the model predictive, green) are also shown for comparison. As in the BIC and AIC, 
  $\max_\Psi \ln L(\Psi;n_{\rm p})$ is offset by its constant part and is multiplied by $-2$.
  Vertical lines and dotted lines of the same colors indicate the optimal $n_{\rm p}$ and the slopes of penalty terms, respectively, for the BIC and AIC. 
  (b) The maximum (blue) and the sample mean (yellow) of 
  $\ln L(\Psi;n_{\rm p})$ at local maxima for each $n_{\rm p}$. For visibility,
  $\ln L(\Psi;n_{\rm p})$ is multiplied by $\sigma^2$ after processed as in Fig.~\ref{fig:likelihoods_lockinginversion}a; the $\sigma^2$ value is here an estimate from our coupling inversion. 
  The sample mean of $\ln L(\Psi;n_{\rm p})$ is evaluated with the sample standard deviation.
  }
  \label{fig:likelihoods_lockinginversion}
\end{figure*}

The optimal estimate of $n_{\rm p}$ is determined by 
the marginal likelihood $L(n_{\rm p})$ of the number of asperities $n_{\rm p}$ (Fig.~\ref{fig:likelihoods_lockinginversion}a). 
$L(n_{\rm p})$ is approximately evaluated by the BIC, which consists of $\max_{\Psi} \ln L(\Psi;n_{\rm p})$ and the penalty on the $n_{\rm p}$ value. 
The maximum of $L(\Psi;n_{\rm p})$ given $n_{\rm p}$ is an increasing function of $n_{\rm p}$ since  increasing the number of bases enables decreasing data residuals.
The slope of $\max_{\Psi} \ln L(\Psi;n_{\rm p})$ accords with that of the BIC penalty at the optimal estimate of $n_{\rm p}$, $n_{\rm p}=5$ for this case. 

The BIC takes 622(.9), 574(.8), 571(.6), 571(.8), 580(.9) for $n_{\rm p}=3,4,5,6,7$, respectively. 
Because the log-likelihood difference 
(the BIC difference times $-1/2$)
is 25 between $n_{\rm p}=3$ and $n_{\rm p}= 5$ and the BIC values for $n_{\rm p}=1,2$ are even larger than that for $n_{\rm p}=3$, we can conclude that $n_{\rm p}>3$ is extremely likely. 
Given the same logic, $n_{\rm p}<7$ is likely with ``five-sigma'' significance. 
Hence, only the cases of $n_{\rm p}=4$, $5$, and $6$ matter in uncertainty evaluations. 
The BIC was almost the same between $n_{\rm p}=5$ and $6$, 
and thus the best model discussion should account for the $n_{\rm p}=6$ case, but the local maxima were almost the same between $n_{\rm p}=5,6$, as seen later in Fig.~\ref{fig:meanlockingcouplingstressing}. 

For comparison, we have also evaluated Akaike's Information Criterion~\citep[AIC;][]{akaike1980use}, which is a commonly used indicator along with the BIC. 
The slope of $\ln L(\Psi;n_{\rm p})$ for $n_{\rm p}>5$ is very close to the $n_{\rm p}$-dependence of the AIC penalty term, hardly constraining the optimal in the sense of AIC (Fig.~\ref{fig:likelihoods_lockinginversion}a). It is in contrast with the parabolic behavior of the BIC derived from $L(n_{\rm p})$. 

The above results show that the likelihood of $n_{\rm p}$ has been a well-behaved unimodal (single-peaked) distribution (Fig.~\ref{fig:likelihoods_lockinginversion}a). Meanwhile, the conditional likelihood $L(\Psi;n_{\rm p})$ of the locking-parameter field $\Psi$ given $n_{\rm p}$ is highly multimodal (multi-peaked) (Fig.~\ref{fig:likelihoods_lockinginversion}b). 
Figure~\ref{fig:likelihoods_lockinginversion}b compares
the maximum of $\ln L(\Psi;n_{\rm p})$ 
with the sample mean of $\ln L(\Psi;n_{\rm p})$ at local maxima for each $n_{\rm p}$. 
The sample mean is evaluated with the sample standard deviation. 
Figure~\ref{fig:likelihoods_lockinginversion}b indicates that the maximum log likelihood is within one standard deviation from the log likelihood averaged over local maxima. 

The results shown in Fig.~\ref{fig:likelihoods_lockinginversion}b suggest that the optimal estimate of our locking inversion might be the best local maximum among the local maxima we found, rather than the true global maximum. 
It sounds reasonable because the estimation of binary variables is a discrete optimization, which generally induces an extreme number of local optima with combinatorial explosions. 
Nonetheless, recalling that an infinitesimal difference in asperity configuration does not affect the discretized locking-parameter fields, we can also perceive that part of multimodality is irrelevant for evaluating well-constrained long-wavelength properties of locking. 
The following analysis supports the latter view. We find that these local optima include one-mesh neighborhoods of the global optimum, which hardly change the likelihood value (i.e., numerically at the global optimum). 

\subsubsection{The cause of the likelihood multimodality and the validity of the optimal estimate}
\label{subsubsec:lockinginv3}

\begin{figure*}
   \includegraphics[width=135mm]{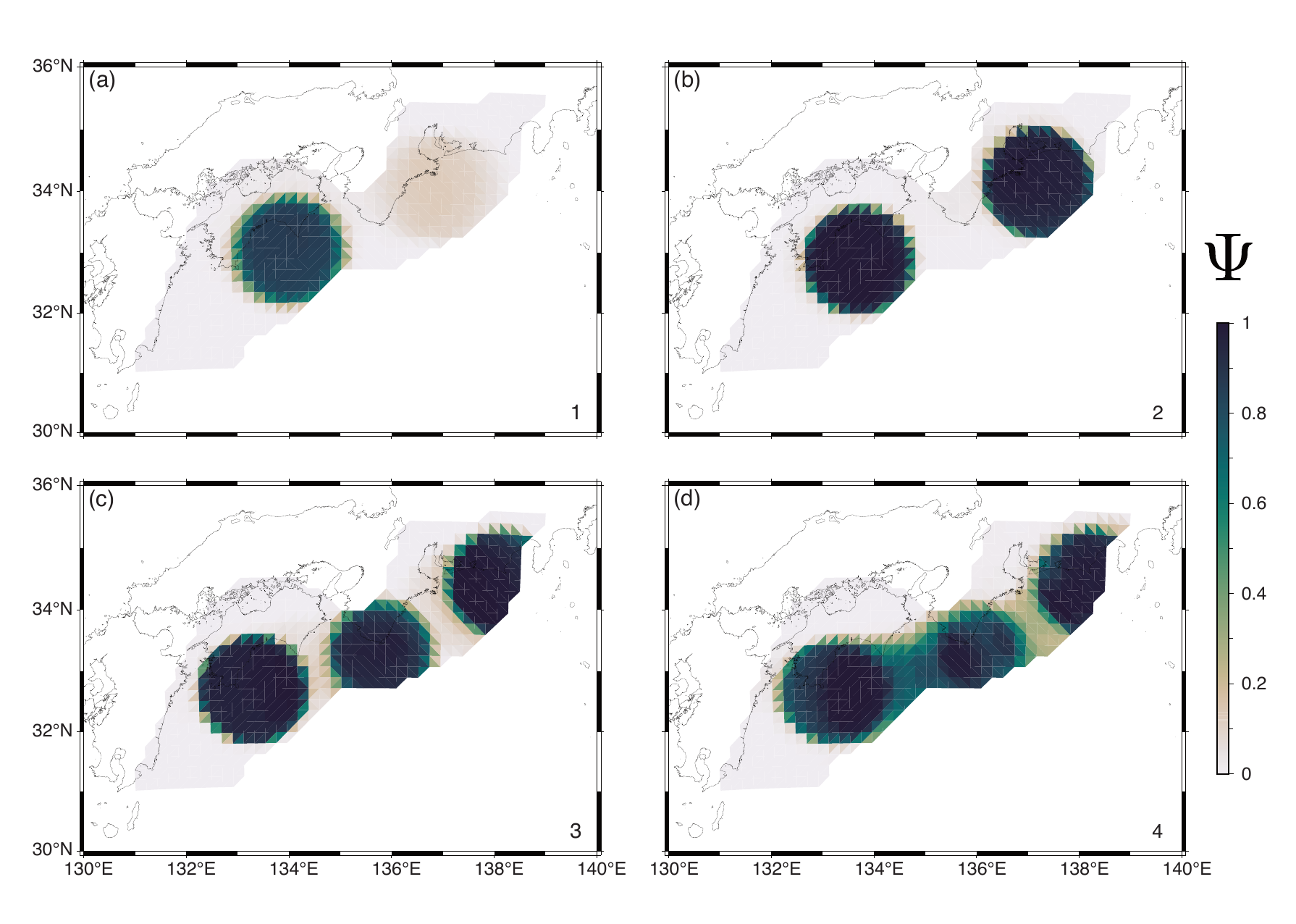}
  \caption{
  The arithmetic means of local optima of the locking-parameter field $\Psi$ for given numbers of asperities $n_{\rm p}=1,2,3,4$. The numbers in panels represent $n_{\rm p}$ values. 
}
  \label{fig:meanlocking}
\end{figure*}

Figure~\ref{fig:meanlocking} shows the sample means of the locking parameter $\Psi$ averaged over the local maxima of $\ln 
L(\Psi;n_{\rm p
})$ for $n_{\rm p}=1$--$4$. 
The quantity evaluated here is not the probability mean and is just a superposition of locally maximum solutions. 
We are aware that this is a very crude approximation of the likelihood mean, but rather, this simplified quantity can clarify the similarity of numerous local maxima. 
For example, the mean locking for $n_{\rm p}=1$ locates either the Nankai area (west) or the Tonankai area (east), with probabilities of around $2/3$ and $1/3$, respectively, indicating the bimodality of $L(\Psi;n_{\rm p
})$ for $n_{\rm p}=1$. 
The mean locking for $n_{\rm p}=2$ locates two asperities on the same locations as the $n_{\rm p}=1$ case but with probability almost 1, meaning the unimodality of $L(\Psi;n_{\rm p})$ for $n_{\rm p}=2$. 
The $n_{\rm p}=3$ case is also effectively unimodal. 
The $n_{\rm p}=4$ case exhibits highly multimodal behaviors, where asperities form a band of the western locking segment from the Cape Shionomisaki to the Bungo-Channel, resulting in a green zone and a beige zone where the mean locking is below 1.

The above behaviors of $n_{\rm p}=1$--$3$ cases 
are relatively simple and can be summarized as follows: 
(i) longer-wavelength patterns are constrained earlier, 
and 
(ii) the multimodality of $L(\Psi;n_{\rm p})$ reflects that there are multiple equal-wavelength features. 
Notice the multimodality prominent in Fig.~\ref{fig:likelihoods_lockinginversion} almost vanishes in the locking-parameter field in Fig.~\ref{fig:meanlocking} for $n_{\rm p}=1$--$3$. 
That is, except for the obvious bimodality of the $n_{\rm p} = 1$ case,
there is only one-mesh-order uncertainty for $n_{\rm p}=1$--$3$ in Fig.~\ref{fig:meanlocking}.  
These one-mesh differences are within numerical errors, so 
$L(\Psi;n_{\rm p
})$ is effectively a sharply peaked distribution with a single peak or double peaks for $n_{\rm p}=1,2,3$ (Fig.~\ref{fig:meanlocking}), not as excessively multimodal as we once imagined from Fig.~\ref{fig:likelihoods_lockinginversion}b. 

\begin{figure*}
   \includegraphics[width=135mm]{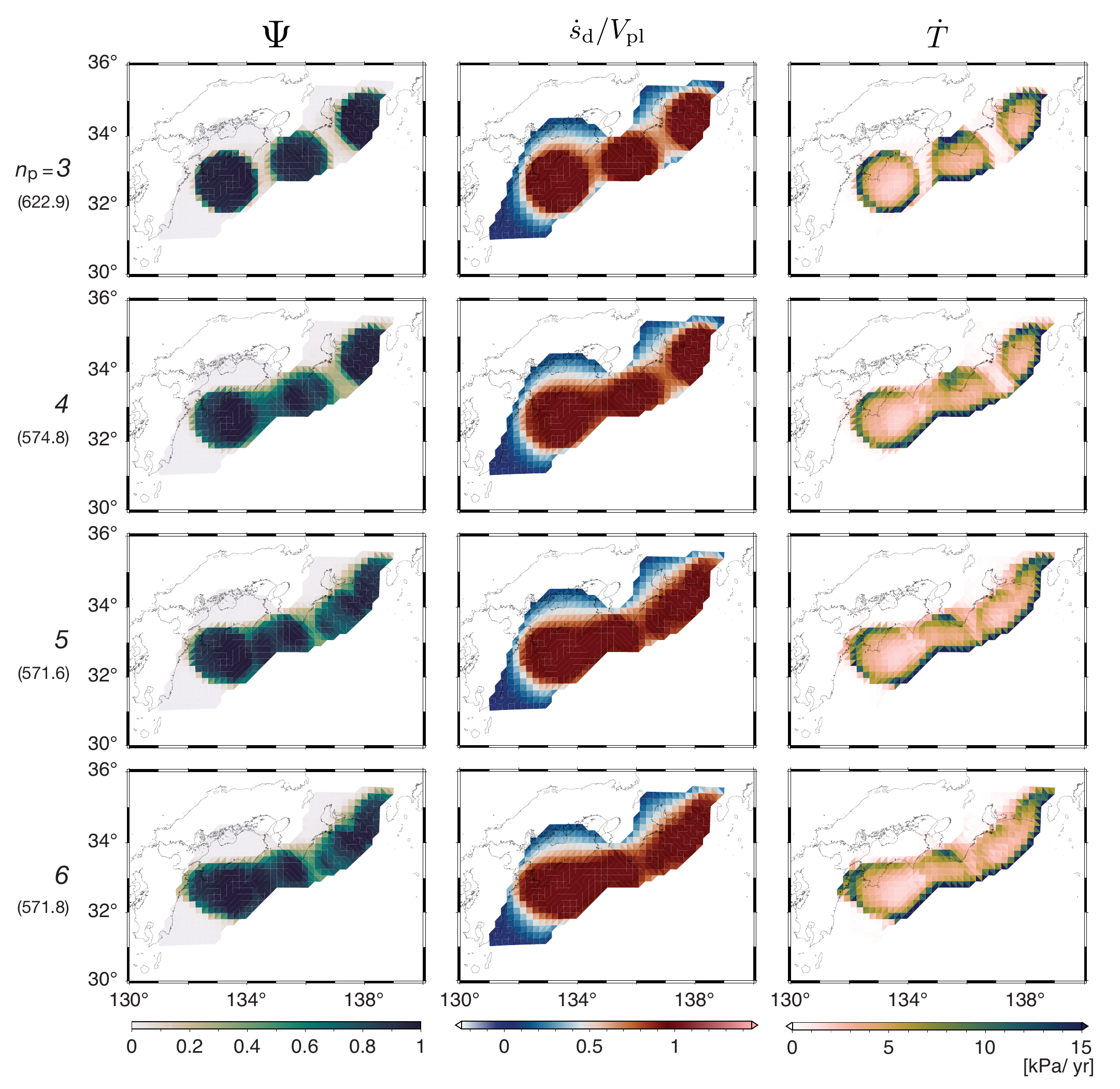}
  \caption{
  The arithmetic means of locally optimum locking $\Psi$ (left), coupling $\dot s_{\rm d}/V_{\rm pl}$ (center), and stressing $\dot T$ (right) fields for given numbers of asperities $n_{\rm p}=3,4,5,6$. The BIC values for respective $n_{\rm p}$ numbers are shown in parentheses for model comparisons.
}
  \label{fig:meanlockingcouplingstressing}
\end{figure*}

In contrast, the multimodality of $L(\Phi;n_{\rm p})$ becomes significant for likely cases $n_{\rm p}=4,5,6$ (Fig.~\ref{fig:meanlockingcouplingstressing}); recall $n_{\rm p}\leq 3\cup n_{\rm p}\geq 7$ is highly unlikely (Fig.~\ref{fig:likelihoods_lockinginversion}a). 
To illustrate their complicated behaviors, we also plot coupling and stressing averaged over local maxima. Those results for $n_{\rm p}=3$ are also plotted for comparison. 

The spatial patterns of plate locking for $n_{\rm p}=4$--$6$ consistently estimate the belts of locked zones with a locking gap separating the western and the eastern. 
In the west belt from around 136${}^\circ$E, the mean locking becomes higher as the number of asperities increases.
The mean locking is almost 1 in this belt for both the best and second-best $n_{\rm p}=5$ and $6$, meaning that most of the sampled local maxima agree with the presence of the western locking belt. 
The eastern belt becomes enlarged along the strike as the number of asperities increases, and the edge exists around 137${}^\circ$E for the more likely cases of $n_{\rm p}=5$ and $6$. 
The locking gap is persistently estimated, supporting our best model, albeit at different locations for $n_{\rm p}=4$ and $n_{\rm p}=5,6$. 
The locking gap is basically estimated at the east of the Cape Shionomisaki: around 136.5${}^\circ$E in the best and second-best cases of $n_{\rm p}=5,6$, and around 137.5${}^\circ$E in the third-best cases of $n_{\rm p}=4$; the $n_{\rm p}=4$ case has 3 higher BIC value (1.5 lower likelihood) than the best case of $n_{\rm p}=5$, so this location variation may be outside the 1.5 standard deviation.

The existence of the locking gap is as above plausible and has been deemed certain from paleoseismicity (\S\ref{subsec:vsTopography}). However, because of the averaging process, the mean locking takes finite values even around the locking gap (Fig.~\ref{fig:meanlockingcouplingstressing}). 
Therefore, just from the mean value, we cannot judge whether the gap location is uncertain or the existence of the locking gap itself is doubtful, 
although the mean stressing visualizes the stress concentration zone despite averaging, supporting the presence of the locking gap.
Opportunely, our asperity-based approach offers a simple way to evaluate the existence of the locking gap. 
We now evaluate the shortest distance of the eastern and western locked segments (segment distance), 
which corresponds to the shortest distance of the easternmost asperity in the western segment and the westernmost asperity in the eastern segment; 
the shortest distance of circular asperities equals their center distance minus the sum of their radii. Non-zero segment distance means the existence of a locking gap.
The non-zero segment distance, namely the segment gap existence, is estimated by 94\% of local optima for the best number of asperities $n_{\rm p}=5$; the mean value (over the local optima) of the segment distance is $30\pm22$ km for $n_{\rm p}=5$. 
The segment gap is estimated to exist by 92\% of local optima for the second best case $n_{\rm p}=6$, which has almost the same BIC as $n_{\rm p}=5$; the mean segment distance is $35\pm30$ km for $n_{\rm p}=6$. 
The segment gap existence is estimated by 100\% of local optima for the third best case $n_{\rm p}=4$ with the mean segment distance of  $60\pm26$ km. 
Considering those results of likely cases $n_{\rm p}=4,5,6$, the $p$-value for the locking gap existence is roughly evaluated below 0.1 but above yet close to 0.05. 
From our model using the current geodetic data, the existence of the locking gap is judged to be fairly statistically significant.

Given these considerations, we conclude that our best estimate of locked zones is consistent with the other local maxima for the likely cases of $n_{\rm p}=4$--$6$, although the estimated locking gap leaves location uncertainty. It indicates that the geodetic observations are able to constrain the asperity configuration of the Nankai subduction zone. 
As such, our model has high precision, and thus we must pay attention to the model bias. 
Figures.~\ref{fig:bestlockingcoupling}, \ref{fig:meanlocking}, and \ref{fig:meanlockingcouplingstressing} insinuate that our transdimensional scheme often sets asperity centers around the shallower part to express the effective ellipticity of the asperities. 
We already noted the lack of data resolution in the shallowest portion near the trough (cf. \ref{subsec:appresult}), but our locking inversion scheme is also not advantageous for resolving the shallowest part. 
We should reemphasize that in our half-space model, the shallowest portion of the half-space is fully coupled outside the meshes, inducing the spurious stress concentration. 
More fundamentally, the shallower locked zone that lies between the ground surface and the locked zone at moderate depth may be almost fully coupled and mispredicted as a shallow extension of the locked zone. Our estimate then may be the worst scenario regarding the locked zone size.
Thorough model improvements are necessary to discuss the shallowest portion just beneath the trough. 


\section{Discussion}
\label{sec:Discussion}

To estimate locked-zone segments, termed asperities in fault mechanics, we have pursued a reduced-order model to describe the locking in the universal sense of friction. 
\citet{wang2004coupling} developed a conceptual classification of kinematic coupling $\dot s_{\rm d}/V_{\rm pl}$ and mechanical couplings, which refer to stressing $\dot T$ and locking $\Psi$ in geodetic inversions thus far although \citet{wang2004coupling} discussed the frictional strength $\Phi$ as well. 
Coupling, stressing, and locking have close but different concepts that characterize the physical properties of faults. 
Among them, plate locking is uniquely a friction-related indicator, 
and thus, its estimation necessarily assumes some frictional boundary condition. 
Previous geodetic locking inversions have formulated locking in the Amontons-Coulomb sense. This study has elucidated that pre-yield and post-yield are the most general definitions of locking and unlocking 
and that interseismic phases reduce many possibilities of friction laws to a single formula of the complementarity of slip and stressing rates, equivalent to the Amontons-Coulomb friction. 
Thus, we reassess the locking inversion using the Amontons-Coulomb friction as the method to infer the locking as the unified status of fault friction rooted in the yield criterion.

This section commits to the comparisons of our results to previous studies. 
The focus of this section is to validate the method, or more widely, the formula of locking, because the scope of this study is the exploration of the formula for locking inversion. A few unexpected findings, such as unlocking below the seismogenic depth and the along-strike locking gap, are also discussed to synthesize them with previous studies. Comparisons with paleoseismicity and structures appear in \S\ref{subsec:vsTopography}. That with slow earthquakes appears in \S\ref{subsec:vsSlowEQs}. 
The limitations of locking inversions and our estimation method are discussed last in \S\ref{subsec:limitations}.

\subsection{Comparison of estimated asperity configuration to historical earthquakes and seafloor topography}
\label{subsec:vsTopography} 
Paleoseismicity along the Nankai subduction zone records that slip zones of megathrust earthquakes are frequently segmented into eastern and western activities~\citep{ishibashi2004status}. 
Teleseismic slip inversions suggest that 
the 1944 Tonankai earthquake and the 1946 Nankai earthquake started around their segmentation boundary, estimating that coseismic slips were small around the segmentation boundary~\citep[e.g.,][]{ichinose2003rupture,murotani2015rupture}. 
The earthquake cycle simulations have predicted that this segmentation boundary corresponds to the locking gap, an unlocked zone between two locked zones, concentrating stress around it and enhancing earthquake nucleation~\citep{kodaira2006cause}. 

\begin{figure*}
\centering
   \includegraphics[width=140mm]{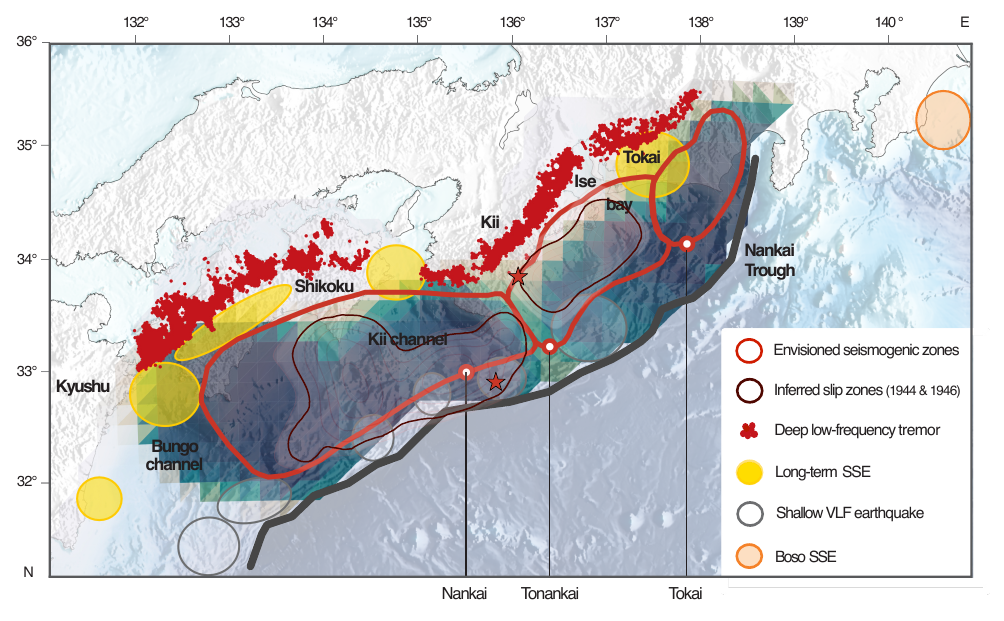}
  \caption{
  Comparison between the estimated  plate locking and seismogenic zones of regular and slow earthquakes. Data compilation by \citet{obara2016connecting} was borrowed for the slow earthquake activities and the envisioned slip zones of megathrust earthquakes. The arithmetic mean of locally optimum locking estimates is overlaid, assuming the optimal number of asperities $n_{\rm p}=5$. For the 1944 Tonankai and 1946 Nankai earthquakes, estimates of rupture initiation points and slip distributions by \citet{kikuchi2003source} and \citet{murotani2015rupture} are also shown. 
}
  \label{fig:vsSSE}
\end{figure*}

Figure~\ref{fig:vsSSE} compares our locking estimate with the envisioned slip zones of the Nankai megathrust earthquakes and the slow earthquake activity~\citep{obara2016connecting}.
For the 1944 Tonankai and 1946 Nankai earthquakes, the estimated slip distributions are borrowed from \citet{kikuchi2003source} and \citet{murotani2015rupture}. The rupture initiation points they assumed are also indicated by stars. 
The 1944 Tonankai earthquake is considered to have caused almost no slip on the west side and a large slip on the east side~\citep{ichinose2003rupture}, though not detailed here. 
For comparison with the point-wise information of rupture initiation points, we now use the arithmetic mean locking of the local optima for $n_{\rm p}=5$, instead of the optimal estimate, to grasp the estimation uncertainty of the locked zone.
The latest decade's findings on slow earthquakes
are not fully reflected in the figure, but we note that slow slip events at shallow depths are found in the Kumano segment around the locking gap we found~\citep{araki2017recurring}.

It should be noted that the location of the locking gap, east of the Cape Shionomisaki, is highly consistent with the estimated slip patterns of the 1944 Tonankai and the 1946 Nankai earthquakes~(Fig.~\ref{fig:vsSSE}).
The rupture initiation point of the 1944 Tonankai earthquake (the right star in Fig.~\ref{fig:vsSSE}) is in the locking gap, and the estimated rupture zone intrudes into the estimated eastern locked segment. 
The rupture initiation point of the 1946 Nankai earthquake (the left star in Fig.~\ref{fig:vsSSE}) is exactly at the eastern edge of the estimated western locked segment, which includes the slip zone of the 1946 Nankai earthquake. 
It is physically natural that the earthquake nucleates at the stress concentration zone~\citep{kodaira2006cause,chen2009scaling}, 
and we were able to extract the associated interseismic behaviors from the surface deformation. 
Although rupture initiation points are unclear on or before 1854, the Cape Shionomisaki has been a segmentation boundary of the eastern and western segments, which have hosted megathrust earthquakes separately~\citep{ishibashi2004status}. 
These facts consistently imply that the locking gap observed from the current geodetic data has been preserved over a geological time scale.

Because it is taken for granted that the seismogenic zones of regular earthquakes are locked~\citep[e.g.,][]{nishikawa2019slow}, the spatial consistency between the previous coseismic slip zone and our locked zone estimate supports that we were able to estimate locking well in our analysis. 
Meanwhile, since the depth of the potential Nankai megathrust earthquakes was only indirectly constrained by other clues such as temperature structures and the seismogenic zones of slow earthquakes, previous indirect assessments of the locked zone would also be validated by our geodetic estimation of locking. 
This may sound tautological, but the estimates of our and previous studies heighten the statistical likelihood of each other.

\begin{figure*}
\centering
   \includegraphics[width=130mm]{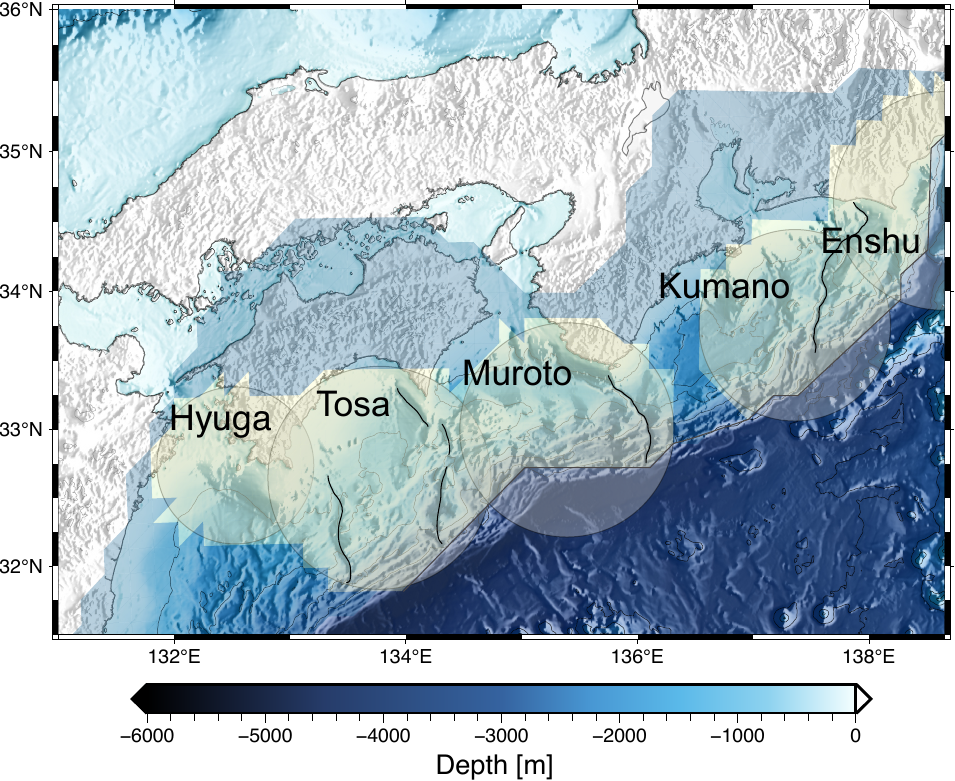}
  \caption{
  Comparison between the optimal configuration estimate of fault-mechanical asperities and seafloor topography. 
  Discretized locked zones (yellow) are plotted on the plate model assumed in inversions (blue, rimed by a solid line), and the estimated circular asperities are overlaid.  
  The seafloor topography is from the IGPP earth relief based on \citet{tozer2019global} with 1000 m contour intervals. 
  Black lines trace submarine canyons and a submarine knoll: from the west, 
  the Aki Canyon, the Ashizuri Canyon (plus neighboring high gradients), the Shionomisaki Canyon, and the Daini-Atsumi Knoll, which separate five basins. 
  The name of the corresponding basin is given to each asperity. 
}
  \label{fig:vsBasin}
\end{figure*}

Next, we look at more detailed features of the asperity configuration. 
\citet{ando1975source} points out that the slip zones of the 1944 Tonankai and 1946 Nankai earthquakes are centered on offshore basins, and this spatial feature is closely investigated by \citet{wells2003basin}. 
By analogy to this spatial correlation between the coseismic slips and offshore basins, 
we compare the estimated locked zone to the seafloor topography~(Fig.~\ref{fig:vsBasin}). 
Interestingly, the estimated five asperities are consistent with offshore basins, including those discussed by \citet{ando1975source}. 
The asperities correlate with five basins: from the west, Hyuga, Tosa, Muroto, Kumano, and Enshu, 
the envisioned rupture segments of the Nankai megathrust earthquake~\citep[e.g.,][]{hirose2022simulation}. 
In terms of paleoseismology, recent analysis suggests six seismogenic segments exist in the Nankai subduction zone~\citep{furumura2011revised}, and our locking inversion now identifies the Enshu segment with the easternmost Omaezaki segment, perhaps because the Omaezaki segment is smaller than our search range of asperity sizes (20 km or larger radii). 
This asperity-topography correspondence becomes clear when the submarine canyons and hills that separate them are shown. 
Asperities in fault mechanics refer to interseismically locked zones, inspired by the term in frictional literature that refers to the topography of frictional surfaces, but far different from the original frictional concept~\citep{scholz2019mechanics}. 
Recalling this chronology, it is interesting that the ``asperity'' in fault mechanics correlates with seafloor topography, the actual surface roughness, but of the earth.

One may notice the Kumano asperity is shifted eastward from the actual Kumano basin in our optimal estimate (Fig.~\ref{fig:vsBasin}). Structural heterogeneities such as stiffness anomaly and a fractured oceanic crust are detected near the edge of the Kumano basin~\citep{kodaira2006cause}, so the actual Kumano asperity, and thus the unlocking gap, may be situated more west. Interestingly, the arithmetic mean of the Kumano asperity is shifted westward from the maximum-likelihood estimate (Figs.~\ref{fig:vsSSE} and \ref{fig:vsBasin}), so more careful locking modeling may mitigate this mismatch of the Kumano basin and asperity.

Existing interpretations on correlations between coseismic slip and geometry let us think that locking may reflect rock property metamorphosis around basins~\citep{wells2003basin} or that frictional resistance of locked zones may cause material deformation and thus mass transfer, resulting in the formation of basins~\citep{song2003large}. 
The locking gap at moderate depth means some frictional anomaly, which could be interpreted as $a>b$ unexpected at the depth/temperature of the locking gap or as fracture energy anomaly of conditional stability, $a\sim b$~\citep{kodaira2006cause}.
However, what is important here is not the cause. 
The key finding for us is that the structures that develop on geological time scales, such as topography, correspond to the frictionally locked zones estimated from the current crustal deformation. 
This correspondence implies that the locked zone, the very candidate of the earthquake source, has been stably preserved rather than randomly varying. 

The five locked zones roughly correspond to the slip zones of past earthquakes: Kumano in 1944, Muroto and Tosa in 1946, Enshu and Kumano in 1854, and all five in 1707~\citep{ishibashi2004status,furumura2011revised}. 
For considering how earthquakes will occur in the future, it will be an important clue that slip profiles of past earthquakes are well explained by the locked zones and their boundary, well correlating with  geological features. Slip zones of megathrust earthquakes in fact change over hundreds of years, but the locked zones capable of hosting them seem as permanent as basins.  

\citet{saito2022mechanically} indicate the highly stressed zone correlates well with basins in their stressing inversion. 
Thus, the correlation between offshore basins and asperities would be plausible, as similar results were obtained with different inversion methods. 
However, the asperity sizes may be debatable, as 
the highly stressed zone in \citet{saito2022mechanically} is significantly narrower than the locked zone in our estimate. 
For example, the eastern segment (Kumano and Enshu in Fig.~\ref{fig:vsBasin}) is further separated into subsegments in the result of
\citet{saito2022mechanically}. 
This asperity size difference between \citet{saito2022mechanically} and our results 
is probably attributed to the difference in the locking and stressing. 
Actually, comparing \citet{saito2022mechanically} and our locking inversion (Fig.~\ref{fig:bestlockingcoupling}), 
the stressing rate estimates are similarly $3$--$6$ kPa/year inside the stressed patches, excluding the artifacts of stress concentration right beneath the trough in our model (\S\ref{subsubsec:lockinginv1}).
As explicated in \S\ref{subsec:semantics}, the stressing rate reveals the rim of the locked zone rather than the locked zone itself (Fig.~\ref{fig:lcs_rel}). 
Because the rim is intrinsically narrower than the body, 
we speculate that \citet{saito2022mechanically} detect the stress-concentrating rims of the locked zones, rather than the giant bodies of the megathrust asperities.



\subsection{Comparison of estimated asperity configuration to slow earthquake activities}
\label{subsec:vsSlowEQs}

Slip zones of regular earthquakes are almost certain to be locked interseismically, but those of slow earthquakes are still under debate. 
The first-order consensus regarding the kinematics of slow earthquakes, at least in the Nankai subduction zone, is that 
their locations are within the transient zone separating the stably creeping zone (no coupling) and the locked zone (full coupling)~\citep{obara2016connecting,baba2020slow}, which is also the case in our results. 
Translated into the RSF, many modelers have read locking and unlocking as the rate-dependence of steady-state friction, the so-called $a - b$ sign, or strictly, the stability of steady sliding affected by the elastic property and fault stress as well as by the frictional properties. 
Possible descriptions of this transient zone
include a mixture of locked and unlocked zones~\citep{ando2012propagation}, 
a broad belt of marginal frictional stability $a\sim b$~\citep{liu2007spontaneous}, 
and an unlocked zone in the stress shadow of the locked zone~\citep{lindsey2021slip}.
Below, we attempt to characterize the interseismic mechanics of these slow earthquakes from our estimates of locking. 

In terms of both width (strike) and depth (dip), the locking estimate overlaps the envisioned slip zones of regular earthquakes, and slow earthquakes at depth are mostly outside it (Fig.~\ref{fig:vsSSE}). 
Documented deep low-frequency tremors are all outside the locked zone estimate.
The estimated locked zones coincide with the fomer focal zones of the same basins~\citep{obara2016connecting} in all basins but the Hyuga. 
When comparing those activities to Fig.~\ref{fig:bestlockingcoupling}b, slow earthquakes at depth occur within the moderately coupled zones. 
Slow earthquakes at shallow depths are complicated, but 
our estimate in this area highly depends on settings other than the data (e.g., priors in coupling inversions, cf., Fig.~\ref{fig:kinematiccouplinginversion}), poorly constrained by observations.

Except for the Hyuga asperity, our results suggest that the seismogenic zones of slow earthquakes are unlocked in long-wavelength and long-time scales. 
There may be some locked zones at short wavelengths, including tremor patches that produce seismic waves~\citep{ando2012propagation}.
However, in terms of the long-wavelength phenomena, such as slow slip events (SSEs), this result has only two possible interpretations: stationary unlocking or apparent unlocking due to the data analysis period. 
The coupling may vary between inter-SSE periods and moments of SSEs~\citep{bartlow2020long,wallace2020slow}.

Thus, we focus on the possible stationarity of unlocking around the slip zones of the slow slip events. 
The locked zone patterns at depth are largely constrained by the onshore data. Around the data period of the onshore data we used (2006/3-2009/12), 
from the west, the Tokai SSEs occur during 2000--2005 and 2013--2015, 
the Kii-Channel SSEs occur during 2000--2002 and 2014--2016, 
the Shikoku SSE occurs during 2005, 
and the Bungo-Channel SSEs occur during 2003 and 2010~\citep{kobayashi2017objective,kobayashi2021application}. 
Because those are not contained in the data analysis period of the onshore data we used, 
although it is a rough discussion, the possibility of the apparent unlocking of the SSE zones is rebutted for the Nankai subduction zone at depth, except for the Hyuga asperity. 

These synthesized results conclude that on geodetic scales, the seismogenic zone of regular earthquakes is locked, whereas the slip zone of slow earthquakes at depth is basically in long wavelength scales coupled but unlocked. 
Then, we move on to its exception, the Hyuga asperity. 
The long-term SSEs occurred in the southern part of the Hyuga basin during 2005--2006, 2007--2008, and 2009--2010~\citep{yarai2013quasi}, slightly overlapping the data analysis period; if this affects the results, rather the Hyuga area should be estimated to be unlocked. 
Namely, time variability is not the cause of the estimated Hyuga asperity.

The Hyuga locked zone includes the slip zones of the 1968 Hyuga-nada earthquake~\citep{yagi1999comparison} and the Bungo-Channel long-term SSEs~\citep{obara2016connecting}. 
Moreover, this zone is supposed to have experienced the fault slip during the 1707 Hoei earthquake~\citep{furumura2011revised}. 
These behaviors of the Hyuga (Bungo-Channel) locked zone are highly complex, but here is one simple, consistent interpretation of those behaviors. 
Namely, the Hyuga locked zone is exceptionally the nucleation zone that often fails to slip faster, as in the Bungo-Channel slow-slip events, but sometimes succeeds, as supposedly in 1707. 
Since this zone has been affected by the model error, probably due to the inland inelastic deformation of Kyushu Island (\ref{subsec:appresult}), the Hyuga locking may be an artifact. 
However, recalling that our inversions estimate the Hyuga asperity after accounting for that model error, we presume that this Hyuga asperity can be the case, although further study is necessary. 

Our results suggest that the slow earthquakes around the Bungo Channel may have a different source process from those of other slow earthquakes. 
Full coupling in an inter-SSE period, similar to the locking of the Bungo Channel SSE slip zone, has been reported in New Zealand~\citep{wallace2020slow} and Cascadia~\citep{bartlow2020long}, 
where the coupling is nearly one during inter-SSE periods while the coupling is zero in total. 
Similar events are reported also in Southern Cascadia~\citep{materna2019dynamically}, 
where a spotty high-coupling zone changes its coupling value repeatedly near the seismogenic zone of Mw$>6.8$ earthquakes, very analogously to the above-mentioned Hyuga locked zone hosting the Bungo-Channel SSEs. 
It is interesting if there are two types of SSEs: one significantly participating in moment release and the other irrelevant in moment evolution.

Last, we compare our results with previous studies that estimate locking. 
\citet{kimura2021mechanical} estimates the locking of the Nankai subduction zone by Bayesian locking inversions first proposed by \citet{johnson2010new}. 
The analysis of \citet{kimura2021mechanical} assumes a belt-shaped locked zone extending along the strike. \citet{sherrill2024locating} employ similar belt-shaped mechanics (discussed in \S\ref{subsec:limitations}) and estimated coupling patterns.
The locking pattern of \citet{kimura2021mechanical} is qualitatively consistent with ours, although the locations of locking-unlocking boundaries are fairly different. 
Examples include that the locked zone around the Bungo-Channel, which is linear and far narrower in \citet{kimura2021mechanical}, while our locking inversion provides a closer coupling pattern to our kinematic coupling inversion. 
Regarding the segment junction of the Tosa and Muroto asperities, where shallow very-low-frequency earthquakes occur,
\citet{kimura2021mechanical} estimates unlocking, and our inversions have excluded meshes through the mesh quality controls (\S\ref{subsec:probset}), thus implicitly assuming unlocking a priori. 
Our estimated locking pattern is rather closer to \citet{sherrill2024locating}, except for the locking-unlocking boundary at depth, where they assumed a different physical constraint. 
Result comparisons are obviously necessary for details, but the scope of this study is to clarify the physics behind locking inversions (\S\ref{sec:2}), and the careful inverse analysis is all future work. 
For now, we trust to our locking estimate, given its consistency with our benchmark solution (\S\ref{subsubsec:lockinginv1}). We expect our solution to be reliable on the locking pattern at depth, where data well constrain the coupling pattern and our consideration of the model errors from elastic Green's functions can improve the estimate (\ref{subsec:appresult}). 

\subsection{Limitations of locking inversions and our results}
\label{subsec:limitations}
Meaningful results have been obtained from a simple model, but the details of locking, unseen in circular asperities, are outside the applicability of our method. 
This subsection summarizes the limitations of the present method to discuss the implications of our inversion results within the method applicability. 

A big assumption of the locking inversion is in neglecting a cohesive zone that separates a locked zone and an unlocked zone. 
We have shown that many friction laws are well represented by the binary of stick and slip (pre-yield and post-yield) within interseismic periods, but albeit an accurate one, it is an approximation. 
An advanced problem is to include a transient unlocked zone ($T=\Phi$ but $\dot \Phi\neq0$) in the locking inversion, as in \citet{sherrill2024locating}. 
For this generalization, another question remains to seek reasonable $\Phi$ evolution. 

Small-scale heterogeneity is also out of scope in this study.  
Our inversion results suggest most of the source regions of slow earthquakes are unlocked, but short-wavelength characters of those zones are outside the scope of our approach. 
Very small patches with sufficiently short recurrence time will satisfy the stress stationarity in the time scale of our interest, so those regions would be detected to be apparently unlocked zones. 
Meanwhile, patches with moderate sizes should be detected even from geodetic observations. 
Those mesoscale locking may be evaluated as the density of the locked zone, which is modeled by \citet{mavrommatis2017physical} as distributed small locked zones. 
The locking density may also be treated in non-binary approaches developed in topology optimization~\citep{ambati2015review}, which have solved similar problems to locking inversions~\citep{eschenauer2001topology}. 

Given those limitations of locking inversions using stick-slip binaries, it is reasonable to question the practical validity of this binary approximation. 
\citet{sherrill2024locating} set 
a transient (unlocked) zone between 
the locked zone $\dot s=0$ and the quasi-steady unlocked zone $\dot T=0$ and estimated the spatial pattern of those trinary phases. 
Even though a specific slip pattern is assumed within the transient zone in their analysis, their results are good touchstones to assess the validity of the binary approximation in the locking inversion. 
For the Nankai subduction zone, 
their results show that the current slip pattern mostly fits the binary view of the conventional locking inversion. 
As long as for the Nankai subduction zone in the interseismic phases, the complementarity of slip and loading rates ($\dot s\dot T\simeq0$) would be a good approximation even practically, although the seismogenic zone of slow earthquakes is sometimes interpreted as a transient zone between the stably creeping zone ($a-b>0$) and locked zone ($a-b<0$)~\citep{liu2007spontaneous,peng2010integrated}. 
\citet{sherrill2024locating} also infer 
transient zones of finite width in Cascadia, so the physical setting of slow earthquake seismogenesis may be tectonics- or timing-dependent. 

We have estimated asperities in the fault mechanical sense, which are defined as locked segments in interseismic periods. 
On the other hand, the term `asperity' refers to an area with a large coseismic slip in strong-motion seismology~\citep{lay1981asperity}.
\citet{das1983breaking,das1986fracture} investigated 
the physical boundary condition for this asperity in the strong-motion seismology, modeling it as a stress-dropping segment surrounded by a constant stress zone~\citep{boatwright1988seismic,irikura2001prediction}. 
This coseismic model of \citet{das1983breaking,das1986fracture} is clearly intended to describe the rupture process of the locked zone. 
Then, according to \citet{das1983breaking} interpretation, 
the asperity in strong-motion seismology will be identified with 
the asperity in fault mechanics conceptually. 
However, our model estimates the envisioned focal zones of the Nankai megathrust earthquakes are totally locked; these locked zones, fault-mechanical asperities, are obviously not the large slip zones for the most recent 1944 Tonankai and 1946 Nankai earthquakes. 
One simple interpretation of this discrepancy is that we might overestimate the locked zone, but the idea of the asperity erosion explored in a series of works~\citep[e.g.,][]{johnson2012challenging,bruhat2017deformation,mavrommatis2017physical} suggests another solution of this contradiction: that is, a locked zone preseismically shrinks~\citep{kato2004interaction}, and thus the interseismic locked zone can be wider than the coseismically unlocked zone. 
Further considerations accounting for realistic earthquake cycles may be necessary for polysemantic asperities to plug geodetically estimated locked zones into strong ground motion assessments.

\remove{Even taking these limitations into account, most of our discussions will remain the same, including the very universal definition of locking and unlocking as the pre- and post-yield phases, asperity-topography correspondence, and arguably, long-term unlocking natures of some slow slip zones.}\add{Although the results of inverse analysis are bound to the data used, the concept of locking holds beyond the observational limit. The number of asperities determined by inverse analysis should not be viewed as the number of asperities that are present, but rather as the limit of resolution that can be objectively detected from the data. Our geodetic detection of megathrust asperities does not rule out the existence of smaller asperities. The fault condition that gradually changes over the interseismic timescale is also reflected in the seismicity level, where repeating earthquakes and microearthquakes operate as creep- and stress-meters of the plate boundary}~\citep{uchida2019repeating,uchide2022stress}\add{. These may facilitate discussion on seismological asperities that exist in hierarchical scales concealed alongside the largest asperities, including asperity-topography correspondence and the long-term unlocking of shallow slow-earthquake zones}~\citep{araki2017recurring}\add{ that trace the locking gap we detected.} As above, one should move to a higher resolution model as data increases, while it seems appropriate to start with a relatively simple model for describing a limited amount of data. 

\section{Conclusion}
Several indicators called mechanical coupling have been proposed to solve the problem of coupling inversions that the coupled zone is always wider than the locked zone. 
The aim of this study is to relate those indicators to the locking in the original sense of friction. 
We organize the frictional physics that locked and unlocked zones follow and start with the very general definition: the locking and the unlocking are defined as the pre-yield and post-yield phases in the yield criterion of the frictional failure. 
Zero slip rate means the locking, and stress at strength means unlocking. 
The condition of locking has been sought as full coupling in the literature, whereas the condition of unlocking has been missed in kinematics. 
The very general definition of locking and unlocking is reduced to a simple formula in the long-term quasi-stationary periods, including interseismic ones, 
which is exactly the physical constraint that has been used in locking inversions: constant slip or constant stress. 
To facilitate the practical use of locking inversion, we also propose a solving scheme implemented as a transdimensional scheme using circular patches, which locates locked segments, termed asperities in fault mechanics. 
\add{We then verify and validate the developed method and the derived formula through actual data analysis.} 
The study area is the Nankai subduction zone in southwestern Japan, and the data are from onshore and offshore geodetic observations.
The optimal estimate concludes that there are five primary asperities consistent with slip zones of historical megathrust earthquakes\add{ separated into the eastern and western belts of locked zones}. 
The spatial distribution of estimated asperities correlates with seafloor topography and locates a segment gap consistent with rupture segmentation documented in paleoseismology, suggesting a direct relationship between intermittent seismicity and persistent geological structures of subduction zones. 
The estimated locked zone mostly does not overlap with slow-earthquake occurrence zones at depth, supporting the hypothesis that the areas hosting slow earthquake clusters are normally in long-term and long-wavelength scales coupled but unlocked. 
However, the Bungo-Channel SSE zone is exceptionally estimated to be locked. 
Given that the Bungo-Channel is thought to be a potential slip zone of the paleoseismic megathrust earthquake, unlike other slow earthquake occurrence zones at depth, the Bungo-Channel may be a locus of earthquake nucleation, which often fails to slip faster but sometimes succeeds. 
Those application results are obviously preliminary\add{ but validate the derived formula of locking by reproducing known features of the regular and slow earthquake activity}. 
The simple, yet ever-present question of what locking is leads to a coherent comprehension of seismological asperities that lock plates during interseismic halts.

\appendix
\section{Conversion of locking-parameter fields to slip deficit fields with 
elementwise-constant discretization}
\label{sec:AppA}
\setcounter{equation}{0}
\setcounter{figure}{0}

The observation equations of the locking inversion consist of
eqs.~(\ref{eq:sdi_obseq}), (\ref{eq:sfromsd}), 
(\ref{eq:slip2traction}), and (\ref{eq:locking_constraint}), 
which, excluding eq.~(\ref{eq:sdi_obseq}), set 
a forward problem to obtain the slip deficit rate from the boundary condition specified by the locking parameter. 
This forward problem is nonlinear, 
but we can find an analytic solution $\dot s_{\rm d}(\Psi)$ of the slip deficit rate $\dot s_{\rm d}$ as a functional of the locking parameter $\Psi$ field~\citep{herman2020locating}.
In this appendix section, we show a simple representation of discretized $\dot s_{\rm d}(\Psi)$ utilizing element sorts.

Let slip deficit rates, locking parameters, traction values, and plate convergence rates of elements be stored in vectors $\dot {\bf s}_d$, $\boldsymbol\Psi$, $\dot {\bf T}$, and ${\bf V}_{\rm pl}$, respectively. 
Now we assume elementwise-constant basis functions of slip rates and locking parameters. 
As in orthodox boundary element models~\citep[e.g.,][]{cochard1994dynamic}, the present study has adopted the center collocation of traction. 
Long-term subduction rates $V_{\rm pl}$ are here assumed to be collocated in the same manner. 
We should note that the center collocation does not converge to the non-subdivision limit within percent order accuracies in three-dimensional problems~\citep{noda2025inconsistency}, unlike two-dimensional cases~\citep[e.g.,][]{sato2020paradox}; nonetheless, the following apply to any forms of the collocation, some of which can overcome this limitation, as elucidated in \citet{noda2025inconsistency}. 

Then, we sort the elements according to the value of $\Psi_i$, 
such that $\boldsymbol\Psi=({\bf 0},{\bf 1})^{\rm T}$, where $^{\rm T}$ denotes transpose. 
After sorting, 
$\dot {\bf s}_d$, $\dot {\bf T}$, and ${\bf V}_{\rm pl}$ are expressed by using their subvectors as
$\dot {\bf s}_{\rm d}=(\dot {\bf s}_{\rm d}^{(0)},\dot {\bf s}_{\rm d}^{(1)})^{\rm T}$, 
$\dot {\bf T}=(\dot {\bf T}^{(0)},\dot {\bf T}^{(1)})^{\rm T}$, 
and $\dot {\bf V}_{\rm pl}=(\dot {\bf V}_{\rm pl}^{(0)},\dot {\bf V}_{\rm pl}^{(1)})^{\rm T}$, 
where the superscripts correspond to the values of $\Psi$.
The discrete traction kernel ${\bf K}$ is also sorted, producing its submatrices ${\bf K}^{({\rm  00})}$, ${\bf K}^{({\rm  01})}$, ${\bf K}^{({\rm  10})}$, and ${\bf K}^{({\rm  11})}$. 
Using these sorted expressions, eq.~(\ref{eq:slip2traction}) becomes
\begin{equation}
\left(
\begin{array}{c}
\dot {\bf T}^{(0)}
\\
\dot {\bf T}^{(1)}
\end{array}
\right)
=
\left(
\begin{array}{cc}
{\bf K}^{({\rm  00})}
&
{\bf K}^{({\rm  01})}
\\
{\bf K}^{({\rm  10})}
&
{\bf K}^{({\rm  11})}
\end{array}
\right)
\left(
\begin{array}{c}
\dot {\bf s}_{\rm d}^{(0)}
\\
\dot {\bf s}_{\rm d}^{(1)}
\end{array}
\right).
\end{equation}
It can be linearly solved for $\dot{\bf s}_{\rm d}$ given $\boldsymbol\Psi$ by using eqs.~(\ref{eq:sfromsd}) and (\ref{eq:locking_constraint}):
\begin{equation}
\dot{\bf s}_{\rm d}(\boldsymbol\Psi)=
\left(
\begin{array}{c}
\dot {\bf s}_{\rm d}^{(0)}
\\
\dot {\bf s}_{\rm d}^{(1)}
\end{array}
\right)
=
\left(
\begin{array}{c}
- {\bf K}^{({\rm  00})-1}{\bf K}^{({\rm  01})}{\bf V}_{\rm pl}^{(1)}
\\
{\bf V}_{\rm pl}^{(1)}
\end{array}
\right).
\label{eq:sdpsi}
\end{equation}
Thus, the conversion of $\Psi$ to $\dot s_{\rm d}$ of eqs.~(\ref{eq:sfromsd}), 
(\ref{eq:slip2traction}), and (\ref{eq:locking_constraint}) is summarized by eq.~(\ref{eq:sdpsi}). 
The locking parameter plays a role in switching the boundary condition imposed to each element, which is a nonlinear but simple routine. 

We then treat the remaining computational implementation. 
Sorting $\dot {\bf s}_{\rm d}$ and other arrays as in eq.~(\ref{eq:sdpsi}) sounds complex in code programming. 
However, because the above procedure is computationally a sub-array extraction based on $\Psi_i$ values, the coding of eq.~(\ref{eq:sdpsi}) is almost a one-liner. In Python, given an $\dot {\bf s}_{\rm d}$ array and a $\boldsymbol\Psi$ array, say, sdr$\_$grid and psi$\_$grid, respectively, $\dot {\bf s}_{\rm d}^{(1)}$ becomes sdr$\_$grid[psi$\_$grid==1], and the set of locked elements is extracted by numpy.where[psi$\_$grid==1]. 
Associated code snippets can be found in the supplement (Open Research Section).

\section{Construction of a benchmark estimate from coupling inversions}
\label{subsec:appresult}
Here, we conduct coupling inversion to construct a benchmark solution of slip-deficit fields, which sets a baseline expected to be reproduced in our locking inversion. 

\subsection{Problem setting of coupling inversions}
The problem setting is basically the same as the locking inversion in the main text (\S\ref{subsec:probset}) except for the prior constraint on the slip deficit. Now we employ a Gaussian distribution and conduct Bayesian coupling inversions. The covariance of the Gaussian prior is weighted by a scale factor, which is an additional hyperparameter of our coupling inversion. Together with the hyperparameters of error statistics ($\sigma^2$ and $\Sigma^2$ in eq.~\ref{eq:errorterm_explicit}), this hyperparameter of the prior distribution is objectively selected by Akaike's Bayesian information criterion~\citep[ABIC, here the same role as model likelihood;][]{akaike1980use,yabuki1992geodetic}. ABIC is evaluated by using Laplace's approximation~\citep{yagi2011introduction}. 

We have employed two Green's functions with different levels of accuracy. One is that of the main text: a rough elastic model, approximating the medium by a half-space homogeneous isotropic Poisson solid, where the fault geometry follows non-planar geometry in the Japan Integrated Velocity Structure Model version 1, while the ground surface of the half-space is approximately set at sea level. The other is an accurate elastic model by \citet{hori2021high}, which is based on the Japan Integrated Velocity Structure Model version 1~\citep{koketsu2009proposal,koketsu2012japan}, accounting for topography, elastic heterogeneity, and the roundness of the earth, as well as the fault geometry.

Three different, popular types of Gaussian priors are employed in this Benchmark analysis. The first one is the Laplacian smoothing prior, which employs squared discrete Laplacian as normalized inverse covariance. The second is traction damping~\citep{saito2022mechanically}, where the logarithm of the prior distribution is proportional to the L2 norm of the traction field. The third one is the roughness constraint of \citet{yabuki1992geodetic}, which imposes the smallness of model parameters (now the slip-deficit rate) at the edge as well as the model-parameter smoothness~\citep{okazaki2021consistent}. 

The smoothing prior in this benchmark test is subtly modified by adding a damping prior of slip deficits at the southwestern edge (element number 0) to calculate concrete ABIC values; while full-ranked prior covariance matrices are required to calculate an absolute value of ABIC, the smoothing prior is rank-deficient in terms of the translational mode of plate boundaries.
This additional constraint will be harmless, since the formulation of slip deficit inversions already removes the rigid-body modes (represented by $V_{\rm pl}$) as almost deformation-free (eq.~\ref{eq:Moment2slipdeficit}). Specifically, we increment the 00-entry, storing the slip-deficit element of the south-western edge, of the above- mentioned discrete Laplacian by 1. We checked that this auxiliary damping constraint at the edge does not change the slip deficit pattern.

\subsection{Results}

\begin{figure*}
    \includegraphics[width=140mm]{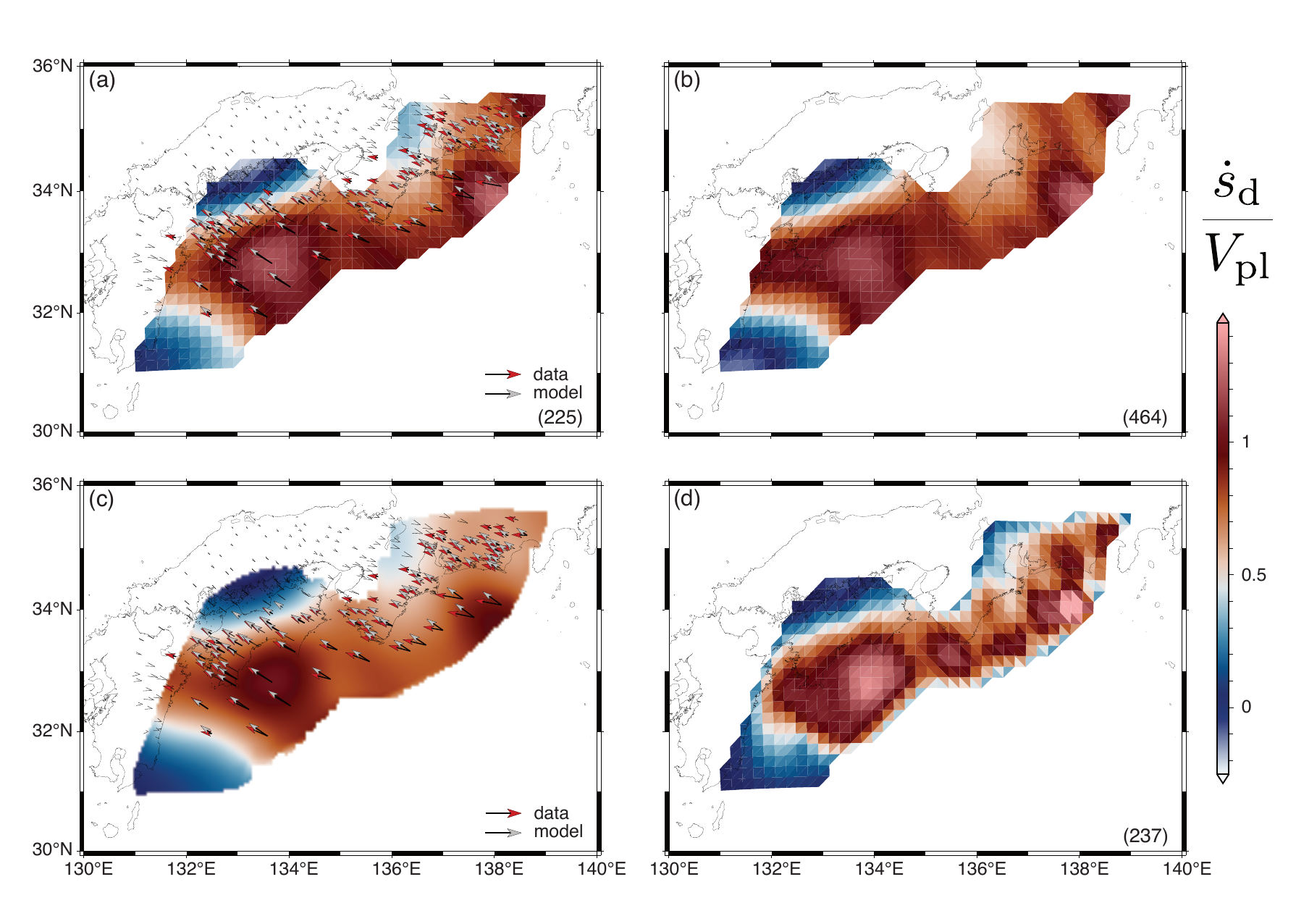}
  \caption{
  Results of coupling inversions. 
  The optimal estimates of coupling ratio $\dot s_{\rm d}/V_{\rm pl}$ are shown for four combinations of prior constraints, Green's functions, and their error considerations. 
  (a) A half-space model with Laplacian smoothing of slip deficits accounting for Green's function errors. 
  (b) A half-space model with Laplacian smoothing of slip deficits without accounting for Green's function errors (a conventional coupling inversion). 
  (c) A realistic elastic earth model of \citet{hori2021high}, in accordance with \citet{koketsu2009proposal,koketsu2012japan}, using slip-deficit Laplacian smoothing and accounting for Green's function errors.
  (d) A half-space model with traction damping, accounting for Green's function errors (a conventional stressing inversion). 
  Data and model surface deformations are shown as arrows for likely models of Fig.~\ref{fig:kinematiccouplinginversion}a and Fig.~\ref{fig:kinematiccouplinginversion}c. The ABIC values of half-space models (a, b, and d) are shown in parentheses for model comparisons, indicating that Fig.~\ref{fig:kinematiccouplinginversion}a is the best solution in our half-space coupling inversions. 
}
  \label{fig:kinematiccouplinginversion}
\end{figure*}

The result of our coupling inversion is summarized in Fig.~\ref{fig:kinematiccouplinginversion}. Four cases are computed to quantify the influence of chosen priors and Green's functions. 

Figure~\ref{fig:kinematiccouplinginversion}a uses the half-space Green's function (the rough model in \S\ref{subsec:probset}) and Laplacian smoothing prior of slip deficits with Green's function errors accounted for~\citep[$\Sigma>0$ in eq.~\ref{eq:errorterm_explicit}, the scale factor of Green's function errors;][]{yagi2011introduction}. 
Figure~\ref{fig:kinematiccouplinginversion}b uses the same setting as Fig.~\ref{fig:kinematiccouplinginversion}a, but without Green's function errors accounted for ($\Sigma\to0$), corresponding to the conventional coupling inversions. 
Figure~\ref{fig:kinematiccouplinginversion}c uses the same settings as that for Fig.~\ref{fig:kinematiccouplinginversion}a, except for the use of high-fidelity Green's function by \citet{hori2021high}, which models realistic topography and elastic structures. Figure~\ref{fig:kinematiccouplinginversion}d uses the same setting as Fig.~\ref{fig:kinematiccouplinginversion}a, 
except for imposing the traction damping prior used in the stressing inversion.
Note that the use of the roughness constraint prior resulted in a similar solution to Fig.~\ref{fig:kinematiccouplinginversion}d but with an inferior ABIC value (i.e., statistically unlikely), thus unplotted now. 

As long as the same Green's function is used
(Fig.~\ref{fig:kinematiccouplinginversion}a, b, and d), the statistical goodness of inversions can be compared by using ABIC (log model likelihood times $-2$, shown in parentheses of Fig.~\ref{fig:kinematiccouplinginversion} panels).
The ABIC values conclude that Fig.~\ref{fig:kinematiccouplinginversion}a is the best estimate for the present half-space setting, and therefore it is our benchmark. 
The associated squared data residual $|{\bf d}-{\bf Hs}_{\rm d}|^2$ was around 10\% of $|{\bf d}|^2$, meaning that the variance reduction was around 90\% in this benchmark. 

Our benchmark solution (Fig.~\ref{fig:kinematiccouplinginversion}a) estimates highly coupled zones consisting of western and eastern sub-regions, mostly consistent with previous coupling inversions~\citep[e.g.,][]{yokota2016seafloor}.
The western one penetrates the deeper portion of the Bungo Channel, suggesting the locked zone at depth near \remove{the }Kyushu Island. 
Meanwhile, 
this benchmark solution estimates shallower portions to be mostly highly coupled. As discussed later in this subsection, however, the shallow-portion coupling largely depends on the prior constraint (Fig.~\ref{fig:kinematiccouplinginversion}a, d), thus poorly constrained by data. 

The influence of model error considerations becomes clear by comparing Fig.~\ref{fig:kinematiccouplinginversion}a and  Fig.~\ref{fig:kinematiccouplinginversion}b.
The conventional models lacking Green's function error considerations (Fig.~\ref{fig:kinematiccouplinginversion}b) indicate a significantly higher ABIC value than that of our benchmark solution (Fig.~\ref{fig:kinematiccouplinginversion}a). 
Consistently, accounting for Green's function errors has mitigated coupling ratios outside $[0,1]$  (Fig.~\ref{fig:kinematiccouplinginversion}a, b), 
which correspond to unphysical subduction faster than $V_{\rm pl}$ or obduction, unreasonable for interseismic plate motions. 
Besides, accounting for Green's function errors moved 
the eastern portion of moderately coupled zones to the shallower side, suggesting that conventional coupling  inversions overestimate the coupling ratio at depth.

The cause of those Green's function errors can be grasped by referring to the inverse analysis using a realistic elastic Green's function (Fig.~\ref{fig:kinematiccouplinginversion}c). 
Figure~\ref{fig:kinematiccouplinginversion}c indicates that 
the half-space model (Fig.~\ref{fig:kinematiccouplinginversion}a) overestimates coupling ratios around the almost fully coupled zones ($\dot s_{\rm d}/V_{\rm pl}\simeq 0.8$) and near the eastern edge.  
Nonetheless, Fig.~\ref{fig:kinematiccouplinginversion}c also shows that the locations of moderately coupled zones ($\dot s_{\rm d}/V_{\rm pl}\simeq 0.5$, white zones in Fig.~\ref{fig:kinematiccouplinginversion}) are not significantly affected. 
Thus, although the absolute value of coupling is debatable, we consider the estimated locations of coupled zones reliable even when using the half-space model, if the Green's function errors are statistically accounted for. 

\subsection{Validation}
Given this positional consistency of moderately coupled zones between the half-space model and the realistic elastic model, the pattern differences in estimated coupling ratios with and without accounting for Green's function errors 
(Fig.~\ref{fig:kinematiccouplinginversion}a, b) would be ascribed to unmodeled inelastic effects. 
The remarkable coupling overestimation of Fig.~\ref{fig:kinematiccouplinginversion}b is actually near the island of Kyushu (left-most coupling) and Itoigawa-Shizuoka Tectonic Line (top right corner), where unmodeled inland inelastic strains exist, and such an error is mitigated by accounting for Green's function error in Fig.~\ref{fig:kinematiccouplinginversion}a. 
The inelastic effect at depth can also come from viscoelasticity because elastic models generally overestimate coupling ratios at depth by dozens of percent due to the overestimation of effective stiffness~\citep{li2015revisiting,li2024revisiting}. Another factor is that the pattern of strain is different for an elastic wedge and an elastic sheet over a viscous wedge~\citep{govers2018geodetic}. Consistently, the eastern segment at depth indicates spurious high coupling ratios near 1 in Fig.~\ref{fig:kinematiccouplinginversion}b, which is removed in the estimates accounting for the errors of elastic Green's functions (Fig.~\ref{fig:kinematiccouplinginversion}a, c). 

The inversion using different priors (Fig.~\ref{fig:kinematiccouplinginversion}a and Fig.~\ref{fig:kinematiccouplinginversion}d) provides a clue to the influence of the prior constraint, as well as a measure of the data resolution.
Figure~\ref{fig:kinematiccouplinginversion}d uses traction damping, the prior constraint of standard stressing inversions. 
Its coupling pattern is consistent with the coupling pattern of previous stressing inversions~\citep{saito2022mechanically}. 
The estimated coupling patterns are similar between slip-deficit smoothing (Fig.~\ref{fig:kinematiccouplinginversion}a) and traction damping (Fig.~\ref{fig:kinematiccouplinginversion}d), but the coupled zones are generally more spotty when using traction damping. 
This comparison reveals that the coupling pattern near the trough largely varies depending on the prior, thus unconstrained by data. 

In summary, in our coupling inversions, the plate coupling is fairly constrained, except around the trough of the subduction zone. 
By explicitly including Green's function errors as error sources, we could detect the effects of inland inelastic deformations, as well as viscoelastic deformations at depth. 
Although the detected inelastic effects require further investigations using physics-based models of inelasticity, it is clear at least that our benchmark solution eliminates evident biases of elastic models, such as uniform high coupling of the eastern area at depth. 
The most striking limitation of our problem setting, clarified through this benchmark, is that the coupling within a few elements from the trough depends on the prior. In our coupling inversion, the trough full-coupling (i.e., locking) is most likely when the plate at moderate depth is fully coupled at the same strike position
(Fig.~\ref{fig:kinematiccouplinginversion}a), but even this result may depend on our assumption of half-space. 
In terms of the prior information, slow earthquakes occur near the trough~\citep{obara2016connecting,araki2017recurring}, implying trough unlocking, while the temperature profile of the Nankai subduction zone suggests $a-b<0$ of the RSF even near the trough~\citep[e.g.,][]{kodaira2006cause}, which will result in trough locking. 
Given these complications, we do not attempt to discuss the shallowest zone with a few elements: the asperity diameter assumed in our application (\S\ref{sec:Application}) is greater than two element intervals.


\section{Likelihood optimizations in multi-asperity locking inversions}
\label{sec:AppC}

Our inversion scheme of locking consists of (process I) the conditional maximum likelihood estimation given a number of asperities and (process II) the comparison of those maximum likelihood estimates according to the BIC. Respective routines include technical topics, which are summarized below. 

The conditional maximum-likelihood search (process I above) of the model parameters in this study is implemented by using the Powell method~\citep{powell1964efficient}. It is a standard direct search method without necessitating the differentiability of the optimization function, which is the log likelihood in this case. Gradient methods such as the BFGS method assuming differentiability of the optimization function often converged to very low likelihood solutions as far as we conducted. Numerical approximations of the Hessian (inverse covariance) matrices perhaps wrongly worked in this scheme. 

The Powell search is not a global search typified by a grid search and thus depends on initial conditions of the optimization. Two different initializations are adopted in this study. One starts with a random asperity configuration. The other sets the initial condition of $n_{\rm p}$ asperities from the best configuration for $n_{\rm p}-1$ asperities plus one randomly generated asperity.

The reasoning of process II is rather technical. The log likelihood function of asperity configuration is remarkably multimodal (Fig.~\ref{fig:likelihoods_lockinginversion}b, discussed in \S\ref{subsubsec:lockinginv3}). 
It complicates the validation of the local search we employed (i.e., not a global search such as a grid search). 
Fortunately, the local maxima provided similar characteristics in long-wavelength scales in our analysis, as investigated in \S\ref{subsubsec:lockinginv3}, and so the uncertainty quantification became a minor topic for our objective. The use of the local search is validated in this sense in the main text.

That multimodality of the asperity likelihood may be parallel to some properties seen in the transdimensional coupling inversions~\citep{tomita2021development}, or the fact that locking inversions treat discrete model parameters with nonlinear equations, generally known to induce multimodality. 
Although it is outside our focus on obtaining a first-order model, there will be many extensions, such as mixed-Gaussian approximations~\citep{ogata1990monte} and replica Markov chain Monte Carlo~\citep{kimura2021mechanical}. 

Additionally, this study sets the search range of model parameters to be a closed space: asperity radii from 20 to 100 km, and asperity centers the bounding box of
31--35.5${}^\circ$N and 131--139${}^\circ$E. 
Those are just for computational tractability, other than the assumption of the minimum radius larger than the mesh interval, which is 20 km roughly equal to or larger than the offshore data point intervals (\S\ref{subsec:probset}). This parameter search simplification did not affect the optimal estimate of the locking. Only the oversimplified cases with $n_{\rm p}=1$--$3$ were affected by the maximum radius limitation of 100 km. 

%



%
%

\section*{Open Research Section}
The GNSS velocity data are available as Table S3 in Supporting Information in \citet{yokota2016seafloor}. 
Python software to implement half-space elastostatic Green's function \citep{nikkhoo2015triangular} is available as ``cutde''~\citep{ben_thompson_2023_8080078}. 
A software package \citep{hori2021high} for elastostatic Green's function and associated fault geometry for the Nankai subduction zone is available as ``Green's Function Library for Subduction Zones''  (\url{https://www.jamstec.go.jp/feat/gflsz/}) from Japan Agency for Marine-Earth Science and Technology (JAMSTEC), which is created by JAMSTEC's own modification of a computer program under development by Earthquake Research Institute, the University of Tokyo. The library includes data modified from Japan Integrated Velocity Structure Model version 1~\citep{koketsu2009proposal,koketsu2012japan} and the Earth Gravitational Model 2008~\citep{pavlis2012development}. 
The code snippets to implement our locking inversions are available in a Zenodo public repository~\citep{sato_2025_15233612}.

\acknowledgments
The authors first greatly appreciate the kind coaching by Roland B\"{u}rgmann concerning this research field and wordings, which have deepened the first author's understanding of the source physics of asperities. 
The authors are deeply grateful to Elizabeth Sherrill, Gareth Funning, Takeshi Iinuma, and Kelin Wang for their insightful comments on the coupling semantics. The authors would also like to thank Eric Lindsey, the anonymous reviewer, the anonymous associate editor, and the editor Anke Friedrich for their careful reviews, which have substantially improved the quality of this research article. This study was supported by JSPS KAKENHI Grant Number 23K19082.

%
%


%
%
%
%
%

\end{document}